\documentclass[aps,reprint,noshowkeys,superscriptaddress]{revtex4-1}
\usepackage{graphicx,dcolumn,bm,xcolor,microtype,multirow,amscd,amsmath,amssymb,amsfonts,physics,longtable,wrapfig,bbold,siunitx,xspace}
\usepackage[version=4]{mhchem}

\usepackage[utf8]{inputenc}
\usepackage[T1]{fontenc}
\usepackage{comment}

\usepackage{hyperref}
\hypersetup{
    colorlinks,
    linkcolor={red!50!black},
    citecolor={red!70!black},
    urlcolor={red!80!black}
}

\usepackage{listings}
\definecolor{codegreen}{rgb}{0.58,0.4,0.2}
\definecolor{codegray}{rgb}{0.5,0.5,0.5}
\definecolor{codepurple}{rgb}{0.25,0.35,0.55}
\definecolor{codeblue}{rgb}{0.30,0.60,0.8}
\definecolor{backcolour}{rgb}{0.98,0.98,0.98}
\definecolor{mygray}{rgb}{0.5,0.5,0.5}

\definecolor{sqred}{rgb}{0.85,0.1,0.1}
\definecolor{sqgreen}{rgb}{0.25,0.65,0.15}
\definecolor{sqorange}{rgb}{0.90,0.50,0.15}
\definecolor{sqblue}{rgb}{0.10,0.3,0.60}

\lstdefinestyle{mystyle}{
    backgroundcolor=\color{backcolour},
    commentstyle=\color{codegreen},
    keywordstyle=\color{codeblue},
    numberstyle=\tiny\color{codegray},
    stringstyle=\color{codepurple},
    basicstyle=\ttfamily\footnotesize,
    breakatwhitespace=false,
    breaklines=true,
    captionpos=b,
    keepspaces=true,
    numbers=left,
    numbersep=5pt,
    numberstyle=\ttfamily\tiny\color{mygray},
    showspaces=false,
    showstringspaces=false,
    showtabs=false,
    tabsize=2
  }

  \usepackage[]{notes2bib}
  \usepackage{siunitx}

  \newcolumntype{d}{D{.}{.}{-1}}

\newcommand{\ie}{\textit{i.e.}\xspace}
\newcommand{\eg}{\textit{e.g.}\xspace}

\newcommand*{\ron}{\rho}

\newcommand{\hH}{\Hat{H}} 
\newcommand{\hh}{\Hat{h}} 
 
\def\md{\mathrm{d}}
\def\bx{{\bf x}}
\def\nm1{{N-1}}
\def\kin{\text{kin}}
\def\cond{\text{cond}}
\def\rr{{\bm{r}}}

\def\ssm{{\smallsetminus}}

\def\s{\text{s}}
\def\resp{\text{resp}}
\def\xc{\text{xc}}
\def\hole{\text{hole}}
\def\h{\text{H}}

\newcommand{\be}{\begin{eqnarray}} 
\newcommand{\ee}{\end{eqnarray}} 

\newcommand{\maX}{\mathbf{Q}} 
\newcommand{\cox}{\mathbf{x}} 

\newcommand{\nucR}{\bm{R}} 
\newcommand{\elr}{\bm{r}} 
\newcommand{\phq}{\bm{q}} 

\newcommand{\TDVP}{\mathbf{A}} 
\newcommand{\TDPES}{\varepsilon}

\usepackage{ulem}
\usepackage{booktabs} 
\usepackage{braket}
\usepackage{mathrsfs}
\usepackage{bm}

\newcommand{\LCT}{Sorbonne Universit\'e, CNRS, LCT UMR 7616, Paris 75005, France}
\newcommand{\MPI}{Max-Planck-Institut f\"ur Physik komplexer Systeme, N\"othnitzer Str. 38, 01187 Dresden, Germany}

\begin{document}	
\title{Exact factorization of a many-body wavefunction beyond the electron-nuclear problem}

\author{Peter \surname{Sch\"urger}}
	\affiliation{\MPI}

\author{Sara \surname{Giarrusso}}
	\email{s.giarrusso1@tue.nl}
	\affiliation{Department of Mathematics and Computer Science \& Institute for Complex Molecular Systems, Eindhoven University of Technology, PO Box, 513 5600 MB Eindhoven, the Netherlands}

\author{Federica \surname{Agostini}}
	\email{federica.agostini@sorbonne-universite.fr}
	\affiliation{\LCT}

\begin{abstract}
This Review is devoted to the presentation of the exact factorization as a framework employed to study a variety of quantum-mechanical many-body problems. Since its original formulation in the 70s, the main applications of the exact factorization were directed towards understanding the properties of multi-component systems, \eg electron-nuclear systems. Especially in the electron-nuclear case, the exact factorization is viewed as an \textsl{exactification} of the Born-Oppenheimer approximation, thus it was, and still is, largely employed in nonadiabatic dynamics. Nonetheless, as early as the 80s, the formalism was employed to study many-electron interacting systems and, quite recently, \ie less than a decade ago, it was extended to study the behavior of molecules in the context in cavity quantum electrodynamics. These formulations, perhaps less popular than the electron-nuclear formulation, have attracted a lot of attention over the years. Therefore, we review here the exact electron-only factorization and the exact photon-electron-nuclear factorization.
\end{abstract}


\maketitle
\section{Introduction}
Solving the \textsl{quantum-mechanical many-body problem} is an extremely challenging and interesting topic that touches various aspects of theoretical chemistry and theoretical physics. The accurate description of the interactions among the fundamental constituents of matter as well as their response to external stimuli is the key to understand the microscopic mechanisms governing the behavior of isolated molecules, biological systems, materials, quantum devices, etc~\cite{Book_GonzalezLindh, Olivucci_CR2066, Caruso_2026, Science_DFT2016, TDDFTbook2018, DMRG_Rev, Birol_ARMR2019, RoadMap2025, Subotnik_S2023}. While the ultimate answer to the many-body problem can be found by solving the Schr\"odinger equation for the (isolated) system of interest, in practice, different theories, hypotheses, and approximations have to be introduced to make the problem manageable and solvable~\cite{Koh-RMP-99, teale2022dft, TDDFTbook2018, Gross_PRA2008, purplebible, BraBurFro-FD-20, Rubio_PNAS2015, Book_GonzalezLindh, DMRG_Rev, Georges_AIP2004, HammesSchiffer_CR2020, Ceperley_RPP2016, Cederbaum_CPL1990, Verstraete_RMP2021, RoadMap2025,Burghardt_PRL2024}. Among the plethora of theories proposed since the origin of quantum mechanics to rationalize the complexity of the many-body problem, we devote this Review to the theory of ``conditional probability amplitudes'', as it was dubbed in 1974 by G. Hunter~\cite{Hunter_IJQC1974,Hunter_IJQC1975_1} in their seminal work on the topic.

The theory of conditional probability amplitudes~\cite{Hunter_IJQC1974,Hunter_IJQC1975_1, Hunter_IJQC1975_2, Bishop_MP1975, Hunter_IJQC1982}, also known as \textsl{exact factorization}~\cite{Gross_PRL2010, Gross_PTRSA2014, Cederbaum_JCP2013, Ghosh_MP2015}, is an overarching strategy in quantum mechanics that allows one to express a general many-body wavefunction as a product of a marginal amplitude and a conditional amplitude.  The marginal amplitude depends only on a selected subset of the system's variables, while the conditional amplitude depends on all variables but is \textsl{parametric} in the marginal variables. Such a decomposition finds its justification in the probabilistic interpretation of quantum mechanics, with the many-body wavefunction yielding a joint probability density which can be expressed as the product of a marginal and of a conditional probability, even though alternative perspectives have been put forth~\cite{Curchod_JCP2025, Burghardt_PRL2024}. The exact factorization was originally introduced for a \textsl{multi-component wavefunction}, dependent on the positions of two sets of distinguishable particles, for instance electrons and nuclei, for which it can be viewed ad an ``exactification'' of the renowned perturbational Born-Oppenheimer approximation~\cite{BO_AP1927, Curchod_FC2025}.

The seminal paper of Hunter did not fully develop the equations allowing to determine the marginal and conditional amplitudes starting from the full time-independent Schr\"odinger equation. However, later work~\cite{Gross_PTRSA2014, Gross_PRL2010, Cederbaum_JCP2013} led to the derivation of two coupled equations: one for the marginal amplitude, in which the conditional amplitude enters through effective scalar and vector potentials, and one for the conditional amplitude, where the marginal amplitude appears in a more intricate, generally non-Hermitian form. Interestingly, the equation for the marginal amplitude is a ``proper'' time-(in)dependent Schr\"odinger equation with an effective Hamiltonian featuring electromagnetic-like gauge potentials, \ie the above-mentioned scalar and vector potentials, allowing one to determine a ``proper'' marginal wavefunction: such a wavefunction yields, in fact, the many-body marginal probability density and its associated current density. While the exact factorization itself provides an interesting viewpoint to analyze a many-body wavefunction, we believe that the derivation of these equations contributed to the success of the theory as a powerful framework to develop original theoretical and computational strategies to solve the many-body problem, thus beyond its application as analysis tool.

The exact factorization of the electron-nuclear \textsl{molecular} problem stimulated much attention in the last decade particularly thanks to the work of Gross and co-workers~\cite{Gross_PRL2010, Gross_JCP2012} that generalized for the first time the theory of conditional probability amplitudes of Hunter to the time domain. Since then, several works have been already devoted to summarize and to give overviews~\cite{EF_bookchapter_2020, TDDFTbook2018, IBELE2024188, Curchod_WIRES2019} or perspectives~\cite{Agostini_PCCP2024} on the implications of the exact factorization in helping our understanding of the many-body electron-nuclear problem, stationary~\cite{Maitra_PRL2020, Gross_PRL2016, Gross_PRL2021, Min_PRL2014, Gross_PTRSA2022} and time-dependent~\cite{Gross_PRL2013, Curchod_JCP2016, Agostini_ADP2015, Agostini_EPJB2021, Agostini_CTC2019, Gross_PRL2017, Gross_PRL2022, Burghardt_PRL2022, Maitra_M2022, Min_JCTC2023, BlumbergerJCP2023, Blumberger_JCP2025, Agostini_PRL2020, Agostini_JCTC2020_1, Gross_PRR2025, Maitra_JCP2019, Min_JCTC2025}. Therefore, this Review will focus on the less conventional, but still rich, body of literature that explored alternative applications of the exact factorization, namely the exact electron-only factorization~\cite{LevPerSah-PRA-84,BuiBaeSni-PRA-89,Gross_PRL2017, GiaGorAgo-CPC-24, GiaAgo-JCP-25} and the exact photon-electron-nuclear factorization~\cite{Agostini_JCP2024_2}. Nonetheless, we find that this Review will greatly benefit from a short and introductory overview of the efforts that several authors have dedicated for more than a decade to the exact electron-nuclear factorization, which we present below. 

In the context of the electron-nuclear \textsl{molecular} problem, the exact factorization has been broadly explored in a variety of physical situations when identifying the marginal wavefunction with the nuclear wavefunction. Some literature on the stationary molecular Schr\"odinger equation has been devoted to the concept of an exact potential energy surface, namely the above-mentioned scalar potential in the effective nuclear Hamiltonian~\cite{Cederbaum_JCP2013, Cederbaum_CP2015, Cederbaum_JCP2014, Sutcliffe_TCA2012, Lefebvre_JCP2015_1, Lefebvre_JCP2015_2, Lefebvre_JCP2016, Lefebvre_MP2017}, and to the appearance of ``spikes'' responsible for interesting node-like features in the nuclear amplitude~\cite{Hunter_IJQC1980, Hunter_IJQC1981, Schild_PRR2021, Wolniewicz_MP1977}. Studies in this direction have been reported in relation to the Born-Oppenheimer (Born-Huang) decomposition of the molecular wavefunction in nonadiabatic conditions~\cite{Curchod_WIRES2019}, to the presence of conical intersections and to the appearance of certain symmetries of the problem~\cite{Jonas_CPL2017, Jonas_JPCA2017, Jonas_JCP2017}. Aiming towards a practical implementation of the stationary exact electron-nuclear factorization, the idea of combining the conditional equation with density-functional theory~\cite{Gross_PRL2016, Gross_JCP2018, Gross_PRB2019} has been exploited as an alternative to approaches like multi-component density functional theory (DFT)~\cite{Gross_PRA2008}. Specifically, the conditional electronic equation has been reformulated by introducing the concepts of conditional electronic and paramagnetic densities, so as to provide a DFT-based framework for the (conditional) electronic problem coupled to the Schr\"odinger-like equation for the (marginal) nuclear problem. In this respect, an alternative -- the first, actually --  strategy to practically solve the conditional equation has been proposed to handle its non-Hermitian component using the optimized effective potential method. The problem of geometric and topological phases arising in molecular systems has been largely investigated employing the perspective of the exact electron-nuclear factorization, both in the stationary and time-dependent formulations. Here, we refer the interested reader to the work of Gross and co-workers~\cite{Min_PRL2014, Requist_PRA2015, Requist_PRA2017}, of Burghardt and Martinazzo~\cite{Burghardt_PRL2024_GP}, of Agostini and co-workers~\cite{Agostini_JPCL2017, Curchod_EPJB2018, Agostini_JPCA2022, Agostini_JPCL2023, Agostini_JPCA2024}, describing how, in general situations, the topological phases in electron-nuclear systems arising at conical intersections revert to geometric properties when analyzed in the eye of the exact factorization. Finally, in its time-dependent formulation, the exact electron-nuclear factorization has been employed extensively in the field of nonadiabatic molecular dynamics~\cite{Min_JPCL2018, Agostini_EPJB2018, Maitra_JPCL2024, Agostini_JCP2024, Min_JCP22, Maitra_PRL2024, Min_TCC2022, Min_JCC2021, Agostini_JCTC2021, Wang_JCP2025}, to simulate relaxation processes of photoexcited molecules~\cite{VM23, Agostini_JCTC2023, Gross_JPCL2017, Tavernelli_EPJB2018, Maitra_PCCP2023, Min_PCCP2019, Agostini_JCTC2020_2, Min_MP2019, Kim_JPCL2018, Choi_CC2019, Kim_JCTC2018, Agostini_MP2024, Agostini_EJPST2023, Maitra_JPCL2022, Min_CP2024, Agostini_JCP2025, Agostini_JCTC2024, Agostini_JCTC2026, Maitra_JCTC2021}, to study the dynamics of molecules under the effect of strong laser fields~\cite{Gross_PRL2010, Suzuki_PCCP2015, Agostini_JCP2021_2, Maitra_PRL2015}, to define atomic masses in molecules~\cite{Scherrer_PRX2017}, to rationalize the emergence of the concept of trajectories in nonadiabatic dynamics~\cite{Gross_EPL2014, Gross_JCP2015, Gross_MP2013, Gross_JCP2014, Gross_PRL2015, Gross_JCTC2016, Franco_JCP2017, Ciccotti_JPCA2020, Ciccotti_EPJB2018, Suzuki_PRA2016, Agostini_JCP2021_1, Agostini_JCP2022, Scribano_JCTC2022, Akimov_JCTC2024}. Further studies have employed the time-dependent exact electron-nuclear factorization to develop a perturbative treatment of nonadiabatic effects~\cite{Schild_JPCA2016, Scherrer_JCP2015}, to draw the connection with the Born-Oppenheimer approximation~\cite{AgostiniEich_JCP2016}, to introduce the idea of time in quantum mechanics~\cite{Schild_PRA2018}, to investigate electronic dynamics in strong fields using the so-called \textsl{inverse} factorization, \ie formulated using a marginal electronic amplitude and a conditional nuclear amplitude~\cite{Suzuki_PRA2014, Maitra_PCCP2017, Maitra_PRL2015}.

In addition to this abundant literature on the exact electron-nuclear factorization, many authors employed the framework of the exact factorization to investigate a variety of many-body problems~\cite{GonZhoRei-EPJB-18, Maitra_PRL2020, Gross_PRL2021,  CohSteGro-PRB-25, Hunter_IJQC1986, Schild_JPCL2021, Ceotto_JCP2025}. Due to their impact in terms of originality and applicability, we focus this Review on the exact electron-only factorization (EEF)~\cite{LevPerSah-PRA-84, Gross_PRL2017, Schild_PRR2020, Schild_PRR2023} and to the exact photon-electron-nuclear factorization (EPENF)~\cite{Maitra_EPJB2018, Tokatly_EPJB2018, Maitra_PRL2019, Maitra_JCP2020, Maitra_JCP2022, Maitra_JCP2021, }. To this end, we organize the remainder of the Review as follows. In Section~\ref{sec: theory} we formulate the exact factorization both in the time-dependent and in the stationary case, adopting a general formalism. Section~\ref{sec:EEF} presents the EEF associated to the stationary Schr\"odinger equation for a many-body system of interacting electrons and how it has been employed in the framework of DFT for developments of functionals. Section~\ref{sec:cavities} is devoted to the extension of the theory to the domain of quantum electrodynamics via the EPENF aiming to describe electron-nuclear dynamics in the regime of strong coupling with photons. Section~\ref{sec:conclusions} concludes the Review.

\section{Exact factorization of a many-body wavefunction}\label{sec: theory}
In this section, we formulate the theory of the exact factorization, assuming a general system of interacting particles. For clarity, we present the time-dependent setting and describe the associated coupled equations for the marginal and conditional amplitudes. The static formalism is recovered by considering a time-independent problem, yielding stationary versions of these coupled equations.

A time-dependent many-body wavefunction $\Psi(\cox,\maX,t)$ representing the state of a system of interacting particles whose positions are indicated as $\maX,\cox$ can be factored as
\begin{align}\label{eqn: EF}
\Psi(\cox,\maX,t) = \chi(\maX,t)\Phi(\cox,t;\maX)
\end{align}
and evolves according to the time-dependent Schr\"odinger equation
\begin{align}\label{eqn: TDSE}
i\hbar \partial_t\Psi(\cox,\maX,t) = \hat H(\cox,\maX)\Psi(\cox,\maX,t)
\end{align}
with Hamiltonian
\begin{align}\label{eqn: H}
\hat H(\cox,\maX) = \hat T_\maX+\hat H_c(\cox,\maX)
\end{align}
We will consider time-independent Hamiltonians, even though time-dependent cases have also been addressed in the literature~\cite{Gross_PRL2010, Gross_JCP2012, Agostini_JCP2021_2, Maitra_PCCP2017, Khosravi2015, Suzuki_PRA2014}. The Hamiltonian of the system, $\hat H(\cox,\maX)$, is the sum of the kinetic energy associated to the marginal variables $\maX$ and of the Hamiltonian $\hat H_c(\cox,\maX)$, which contains the nuclear kinetic energy associated to the conditional ($c$) variables and all interactions. In the electron-nuclear treatment, $\hat T_\maX$ is the nuclear kinetic energy and $\hat H_c(\cox,\maX)$ is the electronic Hamiltonian. 

Inserting the exact factorization~\eqref{eqn: EF} into the time-dependent Schr\"odinger equation~\eqref{eqn: TDSE}, one obtains the evolution equations for the marginal and for the conditional amplitudes~\cite{Gross_PRL2010, Gross_JCP2012, Alonso_JCP2013, Gross_JCP2013},
\begin{align}
i\hbar \partial_t \chi &= \left[\sum_{\nu=1}^{N_n} \frac{[-i\hbar\nabla_\nu+\TDVP_\nu(\maX,t)]^2}{2M_\nu} + \TDPES(\maX,t)\right]\chi \label{eq:marginalMain} \\
i\hbar\partial_t \Phi &= \left[\hat H_{c}(\cox,\maX) + \hat U[\chi,\phi](\maX,t) -\TDPES(\maX,t) \right]\Phi\label{eq:conditionalMain}
\end{align}
The masses of the particles described by $\maX$ are indicated as $M_\nu$ and the index $\nu$ runs of the number of the these particles.

Equations~\eqref{eq:marginalMain} and~\eqref{eq:conditionalMain} have been obtained by imposing the partial normalization condition (PNC)
\begin{align}
    \left\langle \Phi(t;\maX)\big|\Phi(t;\maX)\right\rangle_\cox = \int \left|\Phi(\cox,t;\maX)\right|^2d\cox= 1 \quad \forall\maX,t
\end{align}
stating that the conditional amplitude is normalized at all positions $\maX$ and at all times $t$.

The symbol $\langle \,\,\cdot\,\, \rangle_{\cox}$ stands for an integration over $\cox$, and we use the convention that such variable does not appear explicitly in the term in the braket as it is integrated over. Alternatively, in Sec.~\ref{sec:EEF}, the integration over the conditional variables will be indicated as $\langle \,\,\cdot\,\, \rangle_{\ssm\maX}$, namely an integration over all variables except $\maX$.

The PNC guarantees that $|\Phi(\cox,t;\maX)|^2$ can be identified as a proper conditional probability density, as it is normalized to one, such that the marginal many-body density is given by the squared modulus of the marginal amplitude
\begin{align}
    \left|\chi(\maX,t)\right|^2 = \int \left|\Psi(\cox,\maX,t)\right|^2d\cox
\end{align}
Using this relation, the existence of Eq.~\eqref{eqn: EF} can be easily proven by providing the expressions of the marginal amplitude
\begin{align}
    \chi(\maX,t) = e^{\frac{i}{\hbar}S(\maX,t)}\sqrt{\int \left|\Psi(\cox,\maX,t)\right|^2d\cox}
\end{align}
and of the conditional amplitude
\begin{align}
    \Phi(\cox,t;\maX) = \frac{\Psi(\cox,\maX,t)}{\chi(\maX,t)}
\end{align}
in terms of the full wavefunction $\Psi(\cox,\maX,t)$. The phase $S(\maX,t)$ is so far undertermined, and it is related to the gauge freedom that will be discussed below.

The product form of the full wavefunction $\Psi(\cox,\maX,t)$ is invariant under the phase transformations 
\begin{align}
\tilde\Phi(\cox,t;\maX)&=e^{\frac{i}{\hbar}\theta(\maX,t)}\Phi(\cox,t;\maX)\label{eqn: gauge 1} \\
\tilde\chi(\maX,t)&=e^{-\frac{i}{\hbar}\theta(\maX,t)}\chi(\maX,t)\label{eqn: gauge 2}
\end{align}
Using the fact that
\begin{align}\label{eqn: gauge and g}
    \frac{\tilde\Phi(\cox,t;\maX)}{\Phi(\cox,t;\maX)} = \frac{\chi(\maX,t)}{\tilde\chi(\maX,t)} = g(\maX,t)
\end{align}
since the second term does not depend on $\cox$, and given that the PNC has to hold for both $\Phi(\cox,t;\maX)$ and $\tilde\Phi(\cox,t;\maX)$, one finds that
\begin{align}
\int \left|\tilde\Phi(\cox,t;\maX)\right|^2d\cox = \left|g(\maX,t)\right|^2\int \left|\Phi(\cox,t;\maX)\right|^2d\cox \nonumber\\
= \left|g(\maX,t)\right|^2 = 1
\end{align}
Therefore, from Eqs.~\eqref{eqn: gauge 1},~\eqref{eqn: gauge 2} and~\eqref{eqn: gauge and g} it follows that $g(\maX,t) = e^{\frac{i}{\hbar}\theta(\maX,t)}$, which is the only freedom in the definition of the marginal and conditional amplitudes once the PNC is imposed. Clearly, when solving Eqs.~\eqref{eq:marginalMain} and~\eqref{eq:conditionalMain}, an additional equation has to be imposed to fix the gauge freedom.

As anticipated in the Introduction, the evolution equation determining the marginal amplitude~(\ref{eq:marginalMain}) is an effective time-dependent Schr\"odinger equation, where the coupling to the dynamics of $\Phi(\cox,t;\maX)$ is expressed in terms of a time-dependent vector potential (TDVP)
\begin{align}\label{eqn: TDVP}
\mathbf A_\nu(\maX,t) = \left\langle \Phi(t;\maX)\right|\left. -i\hbar \nabla_\nu\Phi(t;\maX)\right\rangle_{\cox}
\end{align}
and a time-dependent potential energy surface (TDPES)
\begin{align}\label{eqn: TDPES}
\varepsilon(\maX,t) = \left\langle\Phi(t;\maX) \right| \hat H_{c}(\cox,\maX)+\hat U[\Phi,\chi] -i\hbar\partial_t\left| \Phi(t;\maX)\right\rangle_{\cox}
\end{align}
The TDVP encodes information about the momentum field of the particles whose coordinates are given by $\maX$, since 
\begin{align}\label{eq:momentum field}
    \mathbf A_{\nu}(\maX,t) =& \frac{\hbar\text{Im}[\left\langle\Psi(\maX,t)\right|\nabla_\nu
    \left|\Psi(\maX,t)\right\rangle_{\cox}]}{|\chi(\maX,t)|^2} - \nabla_\nu S(\maX,t)        
\end{align}    
where we used the polar representation of the marginal amplitude, $\chi(\maX,t) = e^{\frac{i}{\hbar}S(\maX,t)]}|\chi(\maX,t)|$. The first term on the right-hand side is the marginal momentum field defined from the full wavefunction, whereas the second term is related to the phase of the marginal amplitude. Equation~\eqref{eq:momentum field} shows that the marginal amplitude only carries information about the curl-free component of the marginal momentum field via $S(\maX,t)$, since the curl of a gradient identically vanishes. Therefore, the TDVP carries information about the not-irrotational part. Equation~\eqref{eq:momentum field} also shows the relation between the (so far undetermined) phase of the marginal wavefunction and the gauge freedom, since via a proper choice of $S(\maX,t)$, the curl-free part of the TDVP can be removed.

Under the gauge-like transformations~\eqref{eqn: gauge 1} and~\eqref{eqn: gauge 2}, the TDVP~\eqref{eqn: TDVP} and TDPES~\eqref{eqn: TDPES} transform as standard gauge potentials, namely 
\begin{align}
\tilde{\mathbf A}_\nu(\maX,t)&=\mathbf A_\nu(\maX,t)+\nabla_\nu \theta(\maX,t)\\
\tilde{\varepsilon}(\maX,t)&=\varepsilon(\maX,t)+\partial_t\theta(\maX,t)    
\end{align}
and Eqs.~(\ref{eq:marginalMain}) and~(\ref{eq:conditionalMain}) are form-invariant. In addition, one can easily prove that the TDPES can be written as the sum of a gauge-independent ($GI$) and a gauge-dependent ($GD$) term
\begin{align}\label{eq: tdpes GI+GD}
\TDPES(\maX,t)=\TDPES_{GI}(\maX,t)+\TDPES_{GD}(\maX,t)    
\end{align}
The gauge-independent term contains the contribution arising from $\hat H_c$ (see the first term on the right-hand side of Eq.~\eqref{eqn: TDPES}) and the so-called geometric contribution $geo$ from $\hat U$ (see the second term on the right-hand side of Eq.~\eqref{eqn: TDPES})
\begin{align}
    \TDPES_{GI}(\maX,t) = \TDPES_{c}(\maX,t)+\TDPES_{geo}(\maX,t)
\end{align}
which are defined as
\begin{align}
    \TDPES_{c}(\maX,t)&=\braket{\Phi|\hat H_c|\Phi}_{\cox}\\
    \TDPES_{geo}(\maX,t)&= \sum_\nu \frac{\hbar^2\langle \nabla_\nu \Phi|\nabla_\nu \Phi\rangle_{\cox}}{2M_\nu}-\frac{[\TDVP_\nu(\maX,t)]^2}{2M_\nu}
\end{align}
Instead, the gauge-dependent term is
\begin{align}
    \TDPES_{GD}(\maX,t)&=\braket{\Phi|-i\hbar \partial_t|\Phi}_{\cox}
\end{align}
Equation~(\ref{eq:conditionalMain}) yields the evolution of $\Phi(\cox,t;\maX)$, where the coupling to Eq.~(\ref{eq:marginalMain}) is provided by the coupling operator
\begin{align}
\hat U[\Phi,\chi] = &\sum_\nu \frac{[-i\hbar\nabla_\nu - \mathbf A_\nu(\maX,t)]^2}{2M_\nu}\\
&+\sum_\nu\frac{1}{M_\nu} \left(\frac{-i\hbar \nabla_\nu\chi(\maX,t)}{\chi(\maX,t)}+\mathbf A_\nu(\maX,t)\right)\nonumber\\
&\times\left(-i\hbar \nabla_\nu - \mathbf A_\nu(\maX,t)\right)\nonumber
\end{align}
In the framework of the exact electron-nuclear factorization, a Born-Huang-like representation of the conditional amplitude is often used to draw connections to the Born-Oppenheimer (Born-Huang) perspective. We find interesting to discuss briefly here this point, as it will become useful in Section~\ref{sec:cavities}. The Born-Huang-like expansion of the conditional amplitude reads
\begin{align}\label{eqn: BH in EF}
\Phi(\cox,t;\maX) =\sum_lC_l(\maX,t)\varphi_l(\cox; \maX)
\end{align}
where $\varphi_l(\cox; \maX)$ are the eigenstates of $\hat H _c=\hat H-\hat T_\maX$. The corresponding Born-Huang expansion of the full wavefunction is
\begin{align}\label{eqn: BH in EF}
\Psi(\cox,\maX,t) =\sum_l\chi_l(\maX,t)\varphi_l(\cox; \maX)
\end{align}
such that, by virtue of Eq.~\eqref{eqn: EF}
\begin{align}\label{eqn: coeff BH vs EF}
    \chi_l(\maX,t) = \chi(\maX,t)C_l(\maX,t)
\end{align}
Finally, let us briefly present the exact factorization in the time-independent case. The starting point is the stationary Schr\"odinger equation associated to the Hamiltonian~\eqref{eqn: H}, namely
\begin{align}
    \hat H(\cox,\maX)\Psi_n(\cox,\maX) = E_n\Psi_n(\cox,\maX)
\end{align}
where $\Psi_n(\cox,\maX)$ is the $n$-th eigenstate with associated eigenvalue $E_n$. Each eigenstate of $\hat H(\cox,\maX)$ can be factored as
\begin{align}
  \Psi_n(\cox,\maX)=\chi_n(\maX)\Phi_n(\cox\maX)   
\end{align}
with $\chi_n(\maX)$ and $\Phi_n(\cox;\maX)$ the marginal and conditional time-independent amplitudes. Henceforth, we will remove the subscript $n$ labeling the eigenstates, but all the equations we will show below have to be intended for a given eigenstate of the Hamiltonian (thus to construct the whole set of eigenstates, one should in principle solve the equations below once for every eigenstate).

All the considerations reported above in the time-dependent case are valid also for the stationary problem, namely proof of existence and uniqueness, gauge-transformations and PNC. The equations determining the marginal and conditional amplitudes are, however, slightly different from the time-dependent equations, namely
\begin{align}
\left[\sum_\nu\frac{[-i\hbar\nabla_\nu+\mathbf A_\nu(\maX)]^2}{2M_\nu} + \varepsilon(\maX)\right]\chi &= E\chi \label{eq:marginalMain stationary} \\
\left[\hat H_{c}(\cox,\maX) + \hat U[\chi,\phi](\maX)\right]\Phi&=\varepsilon(\maX) \Phi\label{eq:conditionalMain stationary}
\end{align}
with $E=E_n$. The equation for the marginal amplitude is, once again, an effective Schr\"odinger equation, even though stationary in this case, where the effect of the conditional amplitude is encoded in the exact vector potential $\mathbf A_\nu(\maX)$ and in the exact potential energy surface $\varepsilon(\maX)$, defined similarly to Eqs.\eqref{eqn: TDVP} and~\eqref{eqn: TDPES}. Note, however, that the exact potential energy surface is a gauge-independent quantity since $\tilde{\varepsilon}(\maX)=\varepsilon(\maX)$ while $\tilde{\mathbf A}_\nu(\maX)=\mathbf A_\nu(\maX)+\nabla_\nu \theta(\maX)$. The equation for the conditional amplitude has the form of a Schr\"odinger-like equation, even though the coupling operator $\hat U[\chi,\phi](\maX)$ yields the effect of the marginal amplitude in a non-Hermitian manner, thus preventing from solving Eq.~\eqref{eq:conditionalMain stationary} using standard diagonalization procedures.

In the following sections, we will review how the framework of the exact factorization has been employed in the literature to solve the quantum-mechanical many-body problem for systems of interacting and non-interacting electrons (Sec.~\ref{sec:EEF}), using the formalism of the exact electron-only factorization in the time-independent case, and for a system of interacting electrons, photons and nuclei (Sec.~\ref{sec:cavities}), using the formalism of the exact photon-electron-nuclear factorization in the time-dependent case.

\section{Exact electron-only factorization }\label{sec:EEF}
In the purely electronic setting, the exact factorization formalism has been applied in several distinct ways~\cite{Gross_PRL2017, Maitra_PRL2020, Gross_PRL2021, Schild_PRR2020, Schild_PRR2021, LevPerSah-PRA-84, BuiBaeSni-PRA-89, GiaAgo-JCP-25, GiaGorAgo-CPC-24}.
All the applications discussed in this section treat only the electronic wavefunction and operate within the Born-Oppenheimer approximation.
Because the marginal amplitude in this framework is proportional to the square root of the electron density, the exact electron-only factorization (EEF) provides direct access to density and density-related local potentials. This makes it particularly well suited for analyzing, reconstructing, and modelling the Kohn-Sham potential in density-functional theory (DFT) \cite{KohSha-PR-65, teale2022dft}, an application thread that features prominently in what follows.

Atomic units will be used throughout this section.

\subsection{Asymptotic decay of the electron density}\label{sec:densdecay}
One of the earliest uses of the formalism appears in a paper by by Levy, Perdew, and Sahni \cite{LevPerSah-PRA-84}, where the authors use the conditional amplitude to obtain the asymptotic decay of the electron density.
Their starting point is the time-independent nonrelativistic Schr\"odinger equation
\begin{equation}\label{eq:TISE}
\hH^N (\rr_1, \dots,\rr_N) \Psi_0 (1, \dots, N) = E_0^N \Psi_0 (1, \dots, N)
\end{equation}
where the Hamiltonian is the sum of kinetic energy, electron-electron repulsion and external potential operators (in atomic units), 
\begin{align}
\hH^N (\rr_1, \dots,\rr_N) = \sum_{i=1}^N \left( -\frac{\nabla^2_{\rr_i}}{2} + \sum_{j>i}^N \frac{1}{|\rr_i -\rr_j|}+v (\rr_i)\right)
\end{align}
and $\Psi_0 (1, \dots, N)$ is the corresponding $N$-electron ground state with associated energy $E_0$ and where $i=\sigma_i \rr_i $ indicates the spin and spatial coordinates of the $i$-th electron. Their procedure follows two key steps:
first, decomposing the Hamiltonian into two parts, second, factorizing the electronic wavefunction. 

In the first step, \ie the decomposition of the Hamiltonian, one part encompasses all the operators dealing with the energy of one electron chosen as reference, with position $\rr$; this includes the kinetic energy and the external potential acting on the reference electron, $\hat{h} (\rr) = -\frac{\nabla_\rr^2}{2} + v(\rr)$, and the interaction of the reference electron with all the other electrons $\sum_{i=2}^N \frac{1}{r_{1i}}$ with $r_{1i}=|\rr-\rr_i|$. The other part is simply the remainder: the Hamiltonian for $N-1$ electrons, $\hH ^{N-1}$. Altogether, this reads
\begin{equation}\label{eq:partH}
\hH^N (\rr, \dots,\rr_N)=\hh (\rr)+\sum_{i=2}^N \frac{1}{r_{1i}}+\hH^{N-1}(\rr_2, \dots,\rr_N)
\end{equation}
The second key step, \ie the factorization of the electronic wavefunction, yields
\begin{equation}\label{eq:EFWF}
\Psi_0 (1, \dots, N)=\chi_0(\rr)\Phi_0(\sigma, 2,\dots, N ; \rr)
\end{equation}
where $\chi_0(\rr)$ is identified as the marginal amplitude, while $\Phi_0(\sigma, 2,\dots, N ; \rr)$ as the conditional, which depends \textsl{parametrically} on the coordinate of the reference electron. Note that this coordinate is simply indicated as ``$\rr$'' in both \eqref{eq:partH} and \eqref{eq:EFWF} without the subscript ``$1$''  (and analogously for the spin variable, \ie $\sigma_1=\sigma$).
This notation will remain throughout the section.
The factorization in Eq.~\eqref{eq:EFWF} is constrained by the following condition
\begin{equation}\label{eq:PNC}
\langle \Phi_0(\sigma, 2,\dots, N ; \rr)|  \Phi_0(\sigma, 2,\dots, N ; \rr) \rangle_{\ssm\rr} =1 \quad \forall \rr
\end{equation}
where the Dirac brakets $\langle\dots|\dots \rangle_{\ssm\rr}$ stand for $\int \md \sigma \md 2 \cdots \md N $, \ie integration over all the coordinates of the system but $\rr$. This condition is the PNC introduced in Section~\ref{sec: theory} and allows one to recognize the modulus squared of the marginal as the square root of the electron density, $\ron_0$, (divided by $N$). However, it leaves a phase-factor freedom in the marginal and conditional amplitudes
\begin{align}
\chi_0 (\rr) &=\sqrt{\frac{\ron_0(\rr)}{N}} e^{i S(\rr)} \\
\Phi_0(\sigma, 2,\dots, N ; \rr) &=\frac{\Psi_0(1, \dots, N) e^{- i S(\rr)}}{\sqrt{\ron_0(\rr)/N}}
\end{align}
In Ref.~\cite{LevPerSah-PRA-84}, the gauge choice $S(\rr) = 0 $ is implied and the marginal amplitude becomes a real quantity.

By left-multiplying by $\Phi^*$ and subsequently integrating out the degrees of freedom of $N-1$ electrons (and the spin of the reference electron), they find an effective equation for the square root of the density
\begin{equation}\label{eq:rse}
    \left(-\frac{\nabla_\rr^2}{2} + v_\text{eff}(\rr) \right)\chi_0 (\rr) = \left(E_0^N-E_0^{N-1} \right) \chi_0 (\rr)
\end{equation}
where $\left(E_0^N-E_0^{N-1} \right) = -I_p$ corresponds to the magnitude of the ionization potential of the system and $v_\text{eff}$ is the potential that collects all integrals of the form $\langle \Phi |\hat{o} |\Phi\rangle_{\ssm\rr}$ with $\hat{o}$ one of the operators appearing in Eq. \eqref{eq:partH}.
Equation \eqref{eq:rse} dictates the asymptotic decay
\begin{equation}\label{eq:decay}
    \chi_0 (|\rr| \to \infty) \sim e^{- \sqrt{2\, I_p} |\rr|}
\end{equation}
under the condition that $v_\text{eff} (|\rr| \to \infty) \sim 0$. 
The effective potential $v_\text{eff}$ contains two types of contributions: terms involving the kinetic and Coulomb operators acting on the reference electron, and the term involving the operator $\hat{H}^{N-1}$.
The former vanish at large $|\rr|$ since the conditional amplitude, representing the remaining $N-1$ electrons, becomes increasingly insensitive to the Coulomb potential or variations of the position of the distant reference electron (except possibly along nodal directions, which we do not consider here).
The latter, \ie $\langle \Phi |\hat{H}^{N-1} |\Phi\rangle_{\ssm\rr}$, approaches the ground state energy $E_0^{N-1}$ of the ionized system as $|\rr| \to \infty$, because the conditional amplitude tends to the corresponding ground-state wavefunction. (Only if the electron is removed along a special direction might the conditional amplitude collapse to an excited ionic state.) In Eq. \eqref{eq:rse} this limiting constant, $E_0^{N-1}$, is already subtracted on both sides, yielding the ionization energy on the right-hand side and the exponential decay shown above.

\subsection{Exact decomposition of the Kohn-Sham potential}\label{sec:CAdec}
In a subsequent work by Buijse \textit{et al.} \cite{BuiBaeSni-PRA-89}, the authors establish a connection with the Kohn-Sham (KS) auxiliary system, obtaining an exact decomposition of the exact KS potential in terms of the interacting and non-interacting conditional amplitudes wrapped up into local potentials.
To review their findings, let us start from the definition of the KS Hamiltonian.

If an interacting ground-state density $\ron_0$ can also be generated by a non-interacting Hamiltonian, we refer to that system as the KS system, whose Hamiltonian reads 
\begin{equation}
   \hat{H}^N_\text{s} \left(\rr, \dots, \rr_N \right) = \sum_{i=1}^N \left( -\frac{\nabla^2_{\rr_i}}{2} + v_\s (\rr_i)\right) 
\end{equation}
For the KS system, one can repeat the procedure seen for the interacting case. Namely, first, one partitions the Hamiltonian as
\begin{equation}\label{eq:HSpart}
    \hat{H}_\s^N (\rr, \dots, \rr_N) = -\frac{\nabla^2_\rr}{2} + v_\s(\rr) + \hat{H}^{N-1}_{\text{s} *} (\rr_2, \dots, \rr_N)
\end{equation}
where the symbol $*$ in the subscript of the $N-1$ KS Hamiltonian, $\hat{H}^{N-1}_{\text{s} *}$, specifies that this Hamiltonian contains the external potential $v_\s$ that realizes the $N$-electron density $\ron_0$ \textsl{not} the density of the ionized system; then, one factorizes the KS wavefunction (a Slater determinant), $    \Psi_\text{s}$, in the corresponding marginal and conditional amplitudes
\begin{equation}
    \Psi_\text{s} (1, \dots, N)= \chi_0 (\rr) \Phi_\text{s} \left(\sigma, 2, \dots, N; \rr \right) 
\end{equation}
where the marginal $\chi_0$ is by construction identical to the one in Eq. \eqref{eq:EFWF}.
Since the factorization leads to the same formal structure as in the interacting Hamiltonian case, an effective Schr\"odinger-like equation for the marginal amplitude $\chi_0$ can be obtained in strict analogy with Eq.~\eqref{eq:rse}. However, with the KS Hamiltonian taken as the starting point, the right-hand side of the reduced equation contains the energy difference $E_\text{s}^N - E_{\text{s}*}^{N-1}$, which in turn corresponds to the exact HOMO eigenvalue of KS, $\epsilon_H$. Furthermore, each local potential associated with the terms $\langle \Phi_\text{s} |\hat{o}_\text{s} |\Phi_\text{s}\rangle_{\ssm\rr}$,  where $\hat{o}_\text{s}$ is one of the operators in Eq.~\eqref{eq:HSpart}, also vanishes asymptotically. Together with Eq.~\eqref{eq:decay}, this implies that the exact KS HOMO eigenvalue equals (minus) the ionization energy, $\epsilon_H = - I_p$.
 
In summary, the effective potential $v_\text{eff}$ of Eq.~\eqref{eq:rse} can be expressed in terms of local potentials derived from either interacting conditional amplitudes (left-hand side below) or non-interacting conditional amplitudes (right-hand side):
\begin{equation}\label{eq:veff}
    v + v_\nm1 + v_\kin + v_\cond = v_\s + v_{s, \nm1} + v_{s,\kin} 
\end{equation}
These contributions are defined as
\begin{subequations}
\begin{align}
   & v_\nm1 (\rr) = \langle \Phi |\hat{H}^{N-1} |\Phi\rangle_{\ssm\rr} - E_0^{N-1} \label{eq:nm1}\\
   & v_\kin (\rr) =  \langle\, \left(\nabla_\rr \Phi \right)^2 \rangle_{\ssm\rr} \label{eq:kin} \\
   & v_\cond (\rr) =\sum_{i=2}^N \big{\langle} \frac{\left(\Phi \right)^2}{r1i} \big{\rangle}_{\ssm\rr}
\end{align}
\end{subequations}
with $r1i = |\rr-\rr_i|$.
The KS counterparts $v_{\s,\nm1}$ and $v_{\s,\kin}$ are defined analogously by replacing $\Phi$ by $\Phi_\s$ in Eqs. \eqref{eq:nm1} and \eqref{eq:kin}. Finally, $v$ and $v_\s$ denote the external potentials of the interacting and non-interacting Hamiltonians, respectively.

Equation \eqref{eq:veff} can be inverted to solve for $v_\s$, yielding the decomposition
\begin{equation}\label{eq:vsdec}
    v_\s = v + v_\cond + v_{c, \kin} + v_\resp
\end{equation}
where $v_{c,\kin} = v_\kin - v_{\s, \kin}$ and $v_\resp = v_\nm1 - v_{\s, \nm1}$ represent the correlation-only contributions to the kinetic and $N-1$ potentials. The term $v_\resp$ is commonly referred to as the \textit{response} potential because it can also be expressed through functional derivatives of the 
other potentials. In particular,
\begin{subequations}
\begin{align}
& v_{c,\kin}^\resp (\rr) = \int \ron_0 (\rr') \frac{\delta v_{c, \kin} (\rr')}{\delta \, \ron_0 (\rr)} \md \rr' \\
& v_{\xc-\hole}^\resp (\rr) = \int \ron_0 (\rr') \frac{\delta v_\cond (\rr')}{\delta \, \ron_0 (\rr)} \md \rr' -v_\cond(\rr) \label{eq:xcholeresp}
\end{align}
\end{subequations}
Together, these two response components satisfy
\begin{equation}\label{eq:vresp}
    v_\resp(\mathbf{r}) = v_{c,\kin}^\resp(\mathbf{r}) + v_{\xc-\hole}^\resp(\mathbf{r})
\end{equation}
which is equivalent to the original definition $v_\resp = v_{N-1} - v_{s,N-1}$.

To understand why the second term, $v_{\xc-\hole}^\resp$, is formally more involved than the kinetic response $v_{c,\kin}^\resp$, it is helpful to recall a few facts about the physical meaning of the conditional potential, $v_\cond$.

First, $v_\cond$ is an energy density whose integral yields the  electron-electron interaction energy. Because this interaction is pairwise, a factor $1/2$ must appear
\begin{equation}
    \langle \Psi_0 |\hat{V}_{ee}|\Psi_0 \rangle =\frac{1}{2} \int \ron_0 (\rr) v_\cond(\rr) \md \rr
\end{equation}
with $\hat{V}_{ee}=\sum_{i=1,j>i}^N \frac{1}{|\rr_i -\rr_j|} $.
By contrast the correlation kinetic potential, $v_{c,\kin}$, is also an energy density but it involves a one-body operator. Consequently, no factor 1/2 appears. Multiplied by the electron density and integrated over space, it gives the difference between the interacting and the KS kinetic energy, \ie
\begin{equation}
    \langle \Psi_0 |\hat{T}  |\Psi_0 \rangle -    \langle\Psi_s|\hat{T} |\Psi_s \rangle = \int \ron_0 (\rr) v_{c, \kin} (\rr) \md \rr
\end{equation}
with $\hat{T} = -\sum_{i=1}^N \nabla_{\rr_i}^2/2$.

Second, the conditional potential is usually decomposed into a mean-field contribution -- the Hartree potential --
\begin{equation}
    v_\h (\rr) = \int \frac{\ron_0 (\rr')}{|\rr-\rr'|} \md \rr'
\end{equation}
(note that, by definition, $\int \ron_0 (\rr') \frac{\delta \, v_\h (\rr')}{\delta \, \ron_0 (\rr)}\md \rr' = v_\h (\rr)$),
and a remainder term, which is the electrostatic potential generated by the exchange-correlation (xc) hole,
\begin{equation}
    v_{\xc-\hole}(\rr) = \int \frac{h_\xc(\rr,\rr')}{|\rr-\rr'|}\md \rr'
\end{equation}
The xc hole, $h_\xc$, is a central concept in DFT and is related to the
xc pair-correlation function $g_\xc (\rr, \rr')$ through
\begin{equation}
    h_\xc (\rr, \rr') = \ron_0 (\rr') g_\xc (\rr, \rr')
\end{equation}
 where $g_\xc$ is defined as
\begin{equation}
  g_\xc (\rr, \rr') = \frac{P_2 (\rr, \rr')}{\ron_0(\rr)\ron_0(\rr')}-1
\end{equation}
Here, $P_2 (\rr, \rr')$ is the pair density,
\begingroup
\thinmuskip=0mu \medmuskip=0mu \thickmuskip=0mu
\begin{equation}
    P_2 (\rr, \rr') = N (N-1) \sum_{\sigma, \sigma'} \int |\Psi_0 (\rr \sigma, \rr' \sigma', \bx_3, \dots, \bx_N)|^2 \md \bx_3 \cdots \md \bx_N
\end{equation}
\endgroup
Using $g_\xc$, the xc-hole response potential of Eq. \eqref{eq:xcholeresp} can be rewritten as
\begin{equation}\label{eq:usualxcholeresp}
    v_{\xc-\hole}^\resp (\rr) = \frac{1}{2} \int \frac{\ron_0 (\rr') \ron_0 (\rr'')}{|\rr' - \rr''|} \frac{\delta g_\xc (\rr',\rr'')}{\delta \ron_0 (\rr)} \md \rr' \md \rr''
\end{equation}

\subsection{Structure and physical role of the local potentials}\label{sec:physrole}
\begin{figure}
    \centering
    \includegraphics[width=0.9\linewidth]{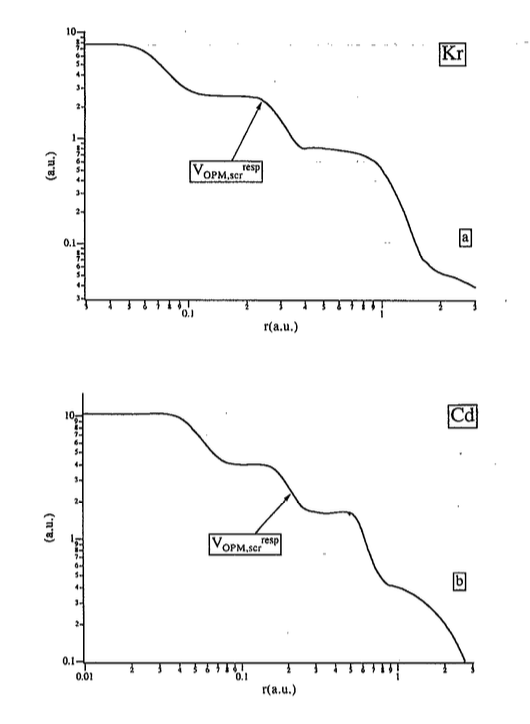}
    \caption{Demonstration of the step character of the potential $v_{\s,N-1}$ for Kr (a) and Cd (b). Reproduced from Gritsenko, O., van Leeuwen, R., \& Baerends, E. J. (1994). The Journal of chemical physics, 101(10), 8955-8963, with the permission of AIP Publishing.}
    \label{fig:shellstructure}
\end{figure}

In subsequent work \cite{GriLeeBae-JCP-94, LeeGriBae-ZPD-95, GriLeeBae-JCP-96, BaeGri-JPCA-97}, the individual components of the KS potential in the decomposition of Eq.~\eqref{eq:vsdec} were evaluated using large configuration-interaction (CI) calculations for the interacting contributions, and either optimized-effective-potential (OEP)-based orbitals \cite{TalSha-PRA-76} or exact orbitals for the KS contributions.
(The exact KS quantities were obtained by inverting the CI density.)
These studies were carried out for selected atoms (Be, Ne, Kr, Cd) \cite{GriLeeBae-JCP-94, LeeGriBae-ZPD-95} and small molecules (XH with X = H, Li, B, F) \cite{GriLeeBae-JCP-96, BaeGri-JPCA-97}. Their value lies in the qualitative insight gained from the systematic patterns that emerge in the structure of the various potential components.
For atoms, the non-interacting part of the response potential, $v_{\s,N-1}$, exhibits a characteristic step or staircase structure, with each atomic shell giving rise to a distinct plateau.
The step structure is clearly visible in Fig.~\ref{fig:shellstructure} where an approximation for $v_{\s, N-1}$ based on OEP calculations is shown for Kr and Cd, taken from Ref. \cite{GriLeeBae-JCP-94}.

This component can be evaluated directly from the (spatial) KS orbitals $\{\varphi_i\}$ that form the KS Slater determinant $\Psi_\s$,
\begin{equation}\label{eq:vsnm1}
    v_{\s,N-1} (\rr)=\sum_{i=1}^\h (\epsilon_H-\epsilon_i)\frac{|\varphi_i (\rr)|^2}{\ron_0 (\rr)}
\end{equation}
This expression rationalizes the staircase structure observed in atoms. Within a region corresponding to a given atomic shell, the density is dominated by a single KS orbital. As a result, the ratio $|\varphi_i|^2/\ron_0$ becomes approximately constant in that region (and equal to one), so $v_{\s,N-1}$ takes the form of a plateau whose height is set by the energy difference $(\epsilon_H-\epsilon_i)$. In this sense, $v_{\s,N-1}$ acts as an effective energy landscape: regions dominated by the $i$th-orbital appear as flat terraces of fixed height. This makes the shell structure stand out much more clearly than when examining the atomic density, its Laplacean, or other commonly used indicators.

At the boundaries between shells, the KS kinetic component $ v_{\s,\kin}$ develops sharp peaks.  These peaks arise because, at such boundaries, the KS conditional amplitude becomes highly sensitive to variations of the position of the reference electron -- precisely the behaviour captured by the gradient in the definition of the kinetic potential [Eq. \eqref{eq:kin}]. The resulting kinetic barriers mark the boundaries between successive plateaus.

The same qualitative behaviour -- plateaus corresponding to regions of constant energy and peaks marking their boundaries -- is also found in the interacting components of the corresponding potentials. The special case of two-electron singlets is particularly illuminating. Here, both $v_{\s, N-1}$ and $v_{\s,\kin}$ vanish identically: there is only one occupied KS orbital (so $\epsilon_i=\epsilon_H$ and therefore $\epsilon_H -\epsilon_i =0$), and the KS conditional amplitude collapses to $\frac{\sqrt{\ron_0 (\rr_2)}}{2}$), which is independent of the reference-electron coordinate. Its gradient therefore vanishes, and so does $v_{\s,\kin}$.
These components are the ones that encode exchange (Fermi statistics) effects arising from the antisymmetry of the KS Slater determinant.
Despite the complete absence of the exchange contributions in $v_{N-1}$ and $v_\kin$ for two-electron singlets, these potentials display the same step-peak pattern seen for $v_{\s, N-1}$ and $v_{\s,\kin}$.
The most striking case where this pattern of the potentials is clearly visible and also very well studied is that of a stretched chemical bond \cite{GriBae-PRA-96, TemMarMai-JCTC-09, GiaNeuBaeGie-JCTC-22}. In this case, the entire structure arises from Coulomb correlation without any contribution from exchange effects.

When a diatomic molecule is stretched, \ie when the internuclear distance $R$ is increased, the KS potential must undergo a dramatic rearrangement in order to reproduce the correct interacting density with a non-interacting system. At equilibrium, most correlation effects are well captured by the xc-hole potential. In this regime, the KS potential remains relatively smooth, and its significant features occur only in regions where the electron density is sizeable. This contrasts sharply with the dissociation limit, where  significant structures develop in regions of extremely low density, as we shall see shortly. As the bond is stretched the two electrons must localize on different atomic fragments to yield the correct dissociation into neutral atoms, each carrying an integer number of electrons.

In the interacting system, this localization arises naturally from the long-ranged Coulomb repulsion, 
which allows one electron to occupy the deeper potential well but prevents the simultaneous localization of both electrons there. In the KS system, however, the electron-electron interaction is absent, so nothing prevents an additional electron to occupy the deeper potential well or an unphysical spreading of the electron density. The only way for the KS system to mimic the correct physics is through a kinetic peak that suppresses the spurious spreading of electron density across the internuclear region,
 and a plateau (or step) that aligns the ionization potentials of the two fragments.
 The kinetic peak acts locally by suppressing the density in the region where it is located.
 
In stretched homonuclear molecules such as H$_2$, where the KS potential does not develop a step, this peak alone prevents the density from spreading into the bond midpoint region, thereby ensuring proper charge localization on each atom. The potential step in a stretched heteronuclear molecule $A - B$ is essential because the KS system always has a single HOMO orbital and therefore a single HOMO eigenvalue $\epsilon_H$, even in the dissociation limit where the two fragments have different physical ionization potentials $I_A$ and $I_B$. To ensure that the KS HOMO simultaneously describes the frontier electrons of both atoms, the KS potential must shift upward locally -- around the more electronegative atom -- so that the effective KS ionization thresholds of the two fragments become equal. This shift is precisely the plateau observed in stretched bonds and it was already theorized by Almbladh and von Barth \cite{AlmBar-INC-85} and Perdew \cite{Per-INC-85}, however it was first attributed to this part of the potential in Ref. \cite{GriBae-PRA-96} and subsequently observed from other accurate calculations of the (numerically) exact KS potential in other model systems \cite{TemMarMai-JCTC-09, KocKraSch-JPCL-21, GiaNeuBaeGie-JCTC-22, KocKraSch-PRR-23}.

\begin{figure}
    \centering
    \includegraphics[width=0.9\linewidth]{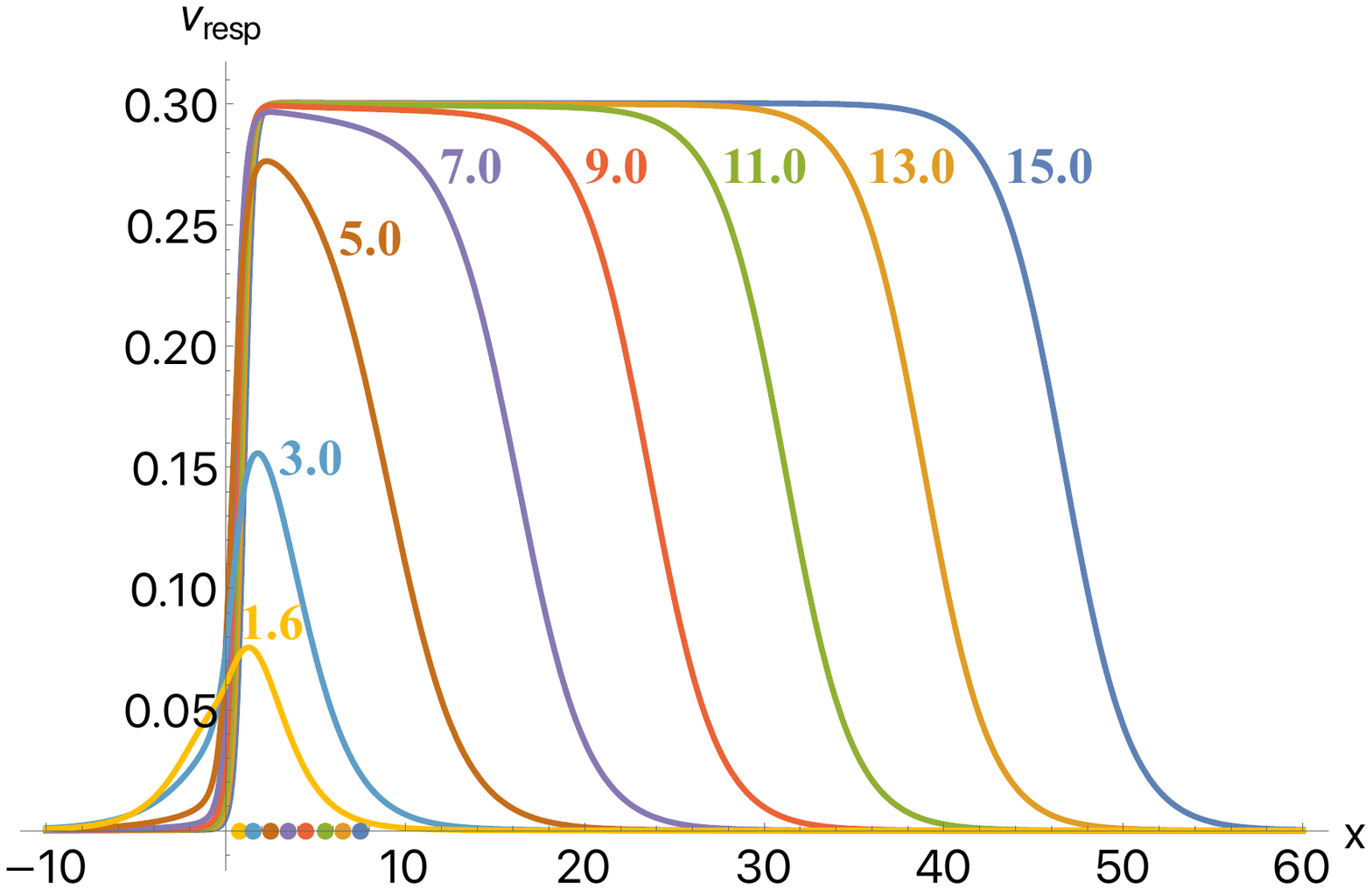}
    \caption{Response potential for the two-electron singlet system of a model diatomic molecule shown at various internuclear distances $R$  (a.u.), labeled on the corresponding curves.
The position of the right nucleus at $R/2$ is indicated by a dot. Adapted from Giarrusso, S., Neugarten, R., Baerends, E. J., \& Giesbertz, K. J. (2022). Journal of chemical theory and computation, 18(8), 4762-4773, with the permission of the American Chemical Society.}
    \label{fig:het_vresp}
\end{figure}

The behavior discussed above is illustrated in Fig.~\ref{fig:het_vresp}, which reports the response potential of a (two-electron) diatomic molecule along the bond axis at different internuclear distances. Note how the structure arises gradually and reaches a definite height ($|I_A-I_B|$) as the dissociation limit is approached, and extends into regions extremely far from the right nucleus (located at $R/2$, marked by a coloured dot), where the density is essentially negligible.
For a more in-depth analysis of the step formation, its spatial extension, and the underlying model, we refer the interested reader to Ref.~\cite{GiaNeuBaeGie-JCTC-22}.

\subsection{Inversions and approximations}
In the development and assessment of density-functional approximations, there is a recurrent need for high-quality KS potentials that can serve as accurate reference data.
Methods aiming at calculating such potentials are usually referred to as (KS) inversion schemes. Several different schemes exist, and the topic is in itself an active field of research \cite{WanPar-PRA-93, LeeBae-PRA-94, JenWas-IJQC-18, CalLatGid-JCP-20, ShiWas-JPCL-21, Gou-JCP-23, KaiKum-JCP-25, HerBakLae-PRB-25}. As a matter of fact, while obtaining the KS density from a given potential is both formally and numerically straightforward, the inverse problem is considerably more challenging. At the root of this difficulty is the fact that, although the Hohenberg-Kohn theorem \cite{HohKoh-PR-64} guarantees a one-to-one mapping between density and potential, small variations in the density may translate into very large variations in the potential \cite{SavColPol-IJQC-03}, making direct inversion extremely sensitive.

Typically, inversion methods rely only on the density (or its response) and discard any additional information that may be available from the underlying electronic-structure calculation, such as the many-electron wavefunction used to obtain the accurate density in the first place.
Instead, an iterative procedure for the reconstruction of the potential that exploits not only the density but also the interacting wavefunction has been developed, based directly on the conditional-amplitude decomposition introduced in Sec.~\ref{sec:CAdec} \cite{RyaKohSta-PRL-15, CueSta-MP-16, CueAyeSta-JCP-15} (see also \cite{ BaeGri-JCP-16, RyaKohCueAyeSta-JCP-16}).
One of the strongest advantages of this procedure is that computing the potential components through the conditional-amplitude integrals circumvents the usual problem that densities expanded in Gaussian bases correspond to KS potentials that oscillate strongly or even diverge \cite{RyaKohSta-PRL-15}.

The success of the exact-factorization-based inversion suggests a broader perspective.
Because the decomposition expresses the KS potential directly in terms of the ingredients that mediate the mapping from the interacting system to its non-interacting counterpart, it exposes the structure that standard explicit density functionals only approximate indirectly.
It is therefore interesting to see whether approximations constructed at the level of these local potential components -- rather than starting from conventional LDA/GGA-type density functionals -- may offer a more direct and potentially more flexible route for overcoming the limitations of current DFT approximations.
In the remainder of this section, we review some works that have explored this alternative strategy.

An early approximation to the response potential -- the so-called GLLB form -- was introduced in Ref.~\cite{GriLeeLenBae-PRA-95}.
Subsequent work has applied the GLLB response potential to obtain accurate band gaps of solids from semilocal functional calculations \cite{KuiOjaEnkRan-PRB-10} and to improve the accuracy of molecular vertical ionization potentials \cite{GriMenBae-JCP-16}.

Another approximation to the response potential was introduced in Ref.~\cite{GiaVucGor-JCTC-18}, based on the strong-interaction or strictly correlated-electrons (SCE) limit of DFT. Because this limit has been combined with the exact electron-only factorization in a number of other works \cite{GiaGorGie-EPJB-18, GroSeiGorGie-PRA-19,GiaGor-JPCA-20, PedCheWhiBur-PRB-22, GiaAgo-JCP-25}, we present it in a nutshell here.

The SCE limit corresponds to the $\lambda\!\to\!\infty$ endpoint of the density-fixed adiabatic connection (AC) and can be formulated as an optimal transport problem~\cite{ButDepGor-PRA-12}. 
Within the AC framework, one considers the family of Hamiltonians
\begin{equation}
\hat{H}_{\lambda}= \hat{T} + \lambda \hat{V}_{ee} + \hat{V}^{\lambda}
\end{equation}
with $\hat{V}^\lambda (\rr_1, \dots,\rr_N) =\sum_{i=1}^N v^\lambda (\rr_i)$ and where the one-body potential $v^\lambda$ is such that all $\hat{H}_\lambda$ share the same ground-state density, equal to the physical density at $\lambda=1$. 
The algebraic manipulations discussed in Secs.~\ref{sec:densdecay} and~\ref{sec:CAdec} can be extended to any $\lambda$ along the AC~\cite{GiaVucGor-JCTC-18}.

In the strong-interaction limit $\lambda\to\infty$, the Hamiltonian has the expansion~\cite{SeiGorSav-PRA-07, GorVigSei-JCTC-09}
\begin{equation}\label{eq:Hsce}
\hat{H}_{\lambda\to\infty}
= \lambda\left(\hat{V}_{ee}+\hat{V}^{SCE}\right) + O(\sqrt{\lambda})
\end{equation}
where $\hat{V}^{SCE}=\sum_{i=1}^N v^{SCE}(\mathbf{r}_i)$ is the one-body potential that minimizes the classical electrostatic energy 
$\hat{V}_{ee}+\hat{V}^{SCE}$ while reproducing the given ground-state density.

A particularly transparent feature of the SCE regime is that the conditional amplitude acquires a simple and exact form. 
Its modulus squared becomes a product of delta functions locating the positions of all other electrons as deterministic functions of the reference-electron coordinate
\begin{equation}\label{eq:phisce}
|\Phi_{\mathrm{SCE}}(\mathbf{r}_2,\dots,\mathbf{r}_N|\mathbf{r})|^2
= \delta(\mathbf{r}_2 - \mathbf{f}_2(\mathbf{r}))\cdots
  \delta(\mathbf{r}_N - \mathbf{f}_N(\mathbf{r}))
\end{equation}
The functions $\mathbf{f}_i(\mathbf{r})$, so-called \textsl{co-motion functions}, are characteristic of the SCE limit. 
They encode how the position of one electron determines those of all others, and are uniquely determined by the density.  
(In principle one should sum over all electron permutations in Eq.~\eqref{eq:phisce}, but this is irrelevant when computing the local potentials, as all terms contribute equally.)

Using Eq.~\eqref{eq:phisce}, the response potential in the SCE limit takes the form
\begin{equation}
v_{\mathrm{SCE},\mathrm{resp}}(\mathbf{r})
= v^{\mathrm{SCE}}(\mathbf{r}) -
  \sum_{i=2}^N \frac{1}{|\mathbf{r} - \mathbf{f}_i(\mathbf{r})|}
\end{equation}
where
\begin{equation}
v^{\mathrm{SCE}}(\mathbf{r})
= \lim_{\lambda\to\infty}\frac{v^\lambda(\mathbf{r})}{\lambda}
\end{equation}
In Ref.~\cite{GiaVucGor-JCTC-18}, the SCE response potential was computed for several spherical atoms. 
It exhibits remnants of atomic shell structure in the form of finite peaks located at radii where the co-motion functions diverge. 
Although this evokes the structure of the exact response potential seen in Fig.~\ref{fig:shellstructure}, the resemblance is limited: instead of broad plateaus separated by narrow boundary peaks (of kinetic origin -- not shown in that figure), the SCE potential consists of a comb-like sequence of localized peaks without extended flat regions.
The same work~\cite{GiaVucGor-JCTC-18} also employed an analytical Heitler--London model to investigate the behaviour of the SCE response potential for a stretched diatomic molecule. 
This Heitler--London \textit{ansatz} was later shown to reproduce many qualitative features of the exact interacting conditional amplitude in stretched bonds~\cite{GiaNeuBaeGie-JCTC-22}, thereby validating its use as a model system for analysing local potential structures in this regime. 
Also in this case, the SCE response potential was found to exhibit remnants of the step-like structure discussed in Sec.~\ref{sec:physrole} and Fig.~\ref{fig:het_vresp}, but with a different scaling behaviour. In particular, adopting a simplified expression for the SCE response potential that becomes exact in the dissociation limit, a follow-up study \cite{GiaGor-JPCA-20}
showed that the height of the SCE step decays as $1/R$ with the internuclear distance $R$.  
By contrast, the exact response potential tends to a constant plateau that aligns the fragment ionization potentials (Fig. \ref{fig:het_vresp}).  
Thus, while the SCE limit captures some structural trends, it fails to reproduce the distinctive scaling behaviour of the exact response potential in bond dissociation, where the plateau height remains fixed as the bond is stretched.

One reason for this discrepancy is that, in the SCE limit, the kinetic contributions are entirely neglected because the kinetic energy operator is formally subleading in the $\lambda\to\infty$ expansion of the Hamiltonian, as per Eq.~\eqref{eq:Hsce}. 
It is worth noting, however, that the SCE minimizer is not a genuine wavefunction but a probability distribution concentrated on a manifold of zero measure, so only $|\Psi_{\text{SCE}}|^2$ is defined.
As a result, if one evaluates the kinetic energy on this distribution, it diverges; a fact that anticipates the difficulties to be tackled when trying to use SCE-based ingredients in expressions involving kinetic quantities.
Even when one proceeds to the next order in $\lambda$, the zero-point-energy (ZPE) correction, and attempts, for example, to approximate the kinetic potential directly via Eq.\eqref{eq:kin}, the resulting ZPE-level potential still provides a picture that is quite far from the qualitative structure expected at the physical coupling strength, $\lambda=1$.
In particular, the ZPE kinetic potential develops a midpoint peak of divergent height in stretched bonds \cite{GroSeiGorGie-PRA-19}.
Although this divergence is mathematically correct within the ZPE order, it bears little resemblance to the finite and well-defined peak around the bond midpoint that we observe in the exact KS potential (Sec. \ref{sec:physrole}).

\begin{figure}
\center
\includegraphics[width=0.81\columnwidth]{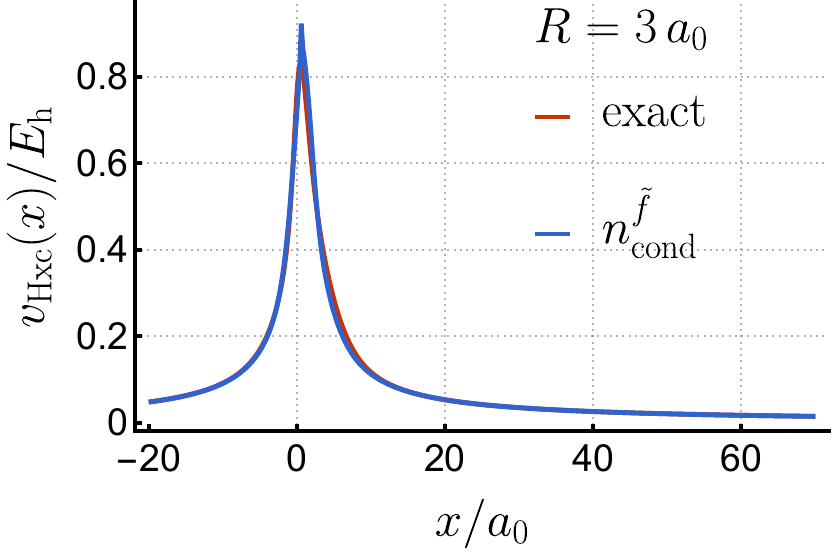}
\includegraphics[width=0.81\columnwidth]{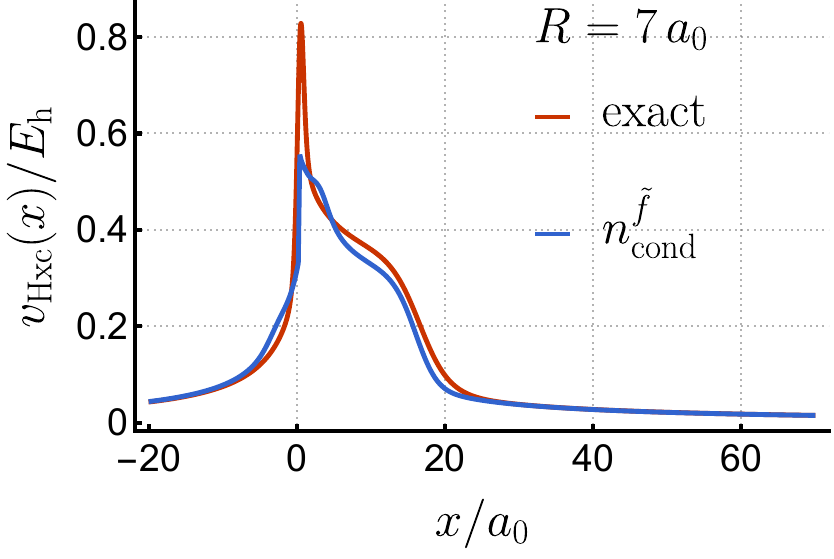}
\includegraphics[width=0.81\columnwidth]{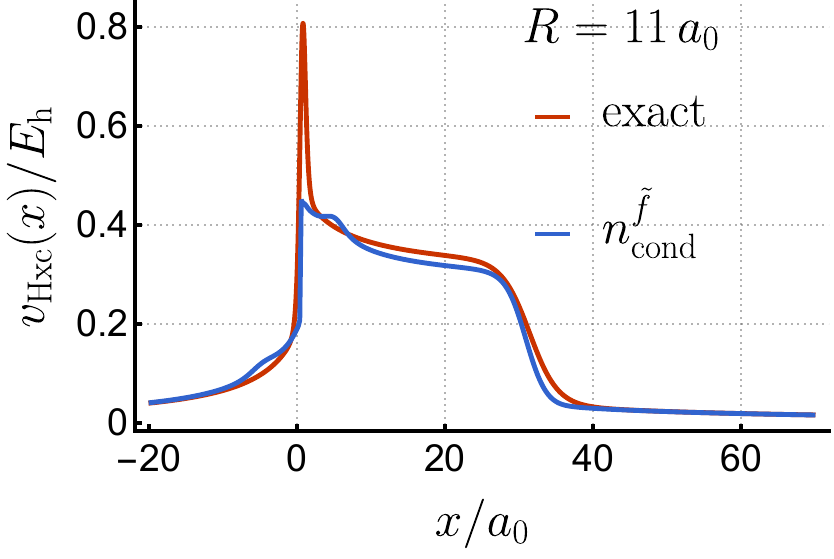}
\caption{Approximate Hxc potential obtained from an approximate conditional density (blue) compared with the exact one (red) for three different internuclear distances: $R = 3\, a_0$ (top panel), $7\, a_0$ (middle panel), and $11\, a_0$ (bottom panel). In the most stretched geometries, the approximate potential accurately reproduces the step structure. Adapted from Giarrusso, S., \& Agostini, F. (2025). The Journal of Chemical Physics, 162(9), licensed under CC BY.}
\label{fig:vhxcfcond}
\end{figure}

A recent work by some of the authors of this Review~\cite{GiaAgo-JCP-25} draws inspiration from the structure of the conditional density in the SCE limit and introduces effective co-motion functions designed to operate at the physical coupling strength $\lambda=1$.
While this construction calls in the strong-interaction limit, its goal is fully aligned with the perspective of the EEF: namely, to model directly the local potentials that emerge from the decomposition discussed in Sec.~\ref{sec:CAdec}, rather than to approximate the energy functional itself. A natural-choice is to model the total Hartree-exchange-correlation potential, $v_{\h\xc}=v_\cond + v_{c,\kin} + v_\resp$, since it collects all the local potentials appearing in the conditional-amplitude decomposition [Eq.~\eqref{eq:vsdec}], apart from the external potential of the interacting problem, which is fixed.
However, the SCE framework is
not directly amenable to construction of these local potentials because it has infinite kinetic energy.
The strategy proposed in Ref.~\cite{GiaAgo-JCP-25} addresses this issue by introducing quantum fluctuations around the classical SCE configuration, thereby regularizing the conditional amplitude and avoiding the pathological kinetic divergence.
When supplied with exact ingredients -- namely, the exact conditional potential used to construct the effective co-motion functions, together with an accurate density -- the method succeeds in reproducing the overall shape of the Hatree-exchange-correlation, $v_{\h\xc}$ across a range of internuclear distances (which corresponds to different correlation regimes), as shown in Fig.~\ref{fig:vhxcfcond}.

To make the approach practical, however, the effective co-motion functions must ultimately be generated from approximate ingredients.
This was explored preliminarily in Ref.~\cite{GiaAgo-JCP-25}, where the SCE functional and the exact-exchange functional were adopted as representative starting points for the strong- and weak-correlation limits, respectively.
The results reveal some limitations: computing approximate co-motion functions is not always inexpensive, and for the SCE functional may not be feasible at all for general systems.
Consequently, further development is needed.
Nonetheless, the proof-of-principle is encouraging 
and opens a new route for constructing approximations to the local potentials arising from the EEF.

\subsection{Beyond ground states}
So far we have remained confined to ground-state theory and time-independent phenomena; however, the EEF has also been extended beyond ground states.
The first application of this formalism to the time-dependent Schr\"odinger equation for the electronic wavefunction was presented in Ref.~\cite{Gross_PRL2017}. There, the authors derived the time-dependent equation for the electronic marginal amplitude, which closely mirrors the structure of the electro-nuclear factorization, most notably through the appearance of a time-dependent scalar potential and a time-dependent vector potential,
\begin{equation}\label{eq:TDEEF}
 i \partial_t \chi (\rr, t) = \left( \frac{[-i \nabla_\rr +\mathbf{A}(\rr,t)]^2}{2}  +  \varepsilon (\rr, t)\right)\chi (\rr, t) 
\end{equation}
In Ref.~\cite{Gross_PRL2017}, the numerically exact dynamics of one-dimensional many-electron model systems subject to a time-dependent laser field were computed, including two- and three-electron systems resembling helium and lithium, respectively, with the initial state taken to be the corresponding electronic ground state in each case.  
The exact time-dependent scalar potential was compared with that obtained within the time-independent conditional amplitude (TICA) approximation, in which only the marginal amplitude is propagated in time while the conditional amplitude is kept fixed.
Under these conditions, TICA was found to provide a reasonable description for the two-electron system, while its accuracy deteriorated for three electrons, indicating the increasing importance of the time dependence of the conditional amplitude as the complexity of electronic correlations grows. 

The analysis of the TICA approximation was subsequently extended in Ref.~\cite{Schild_PRR2020} (see also Ref.~\cite{kocak2022eef}), where the dynamics of a model system with an initial state resembling the lowest triplet state of helium was investigated.
In this case, TICA was found to perform satisfactorily at high laser frequencies but to fail at lower frequencies, even for two-electron cases, further underscoring the importance of accounting for the time dependence of the conditional amplitude for more challenging cases of electron correlation effects.

Although several studies \cite{Gross_PRL2017, Schild_PRR2020, KocKraSch-JPCL-21, KocKraSch-PRR-23} already anticipated the theoretical importance of the electronic vector potential, all simulations were restricted to one-dimensional systems.
As a consequence, the role of the vector potential -- whose gauge-independent effects related to its not-irrotational component are inherently tied to spatial dimensionality -- could not be explicitly analysed. 
This gap was partially addressed in Ref.~\cite{GiaGorAgo-CPC-24}, where the EEF was applied to general stationary solutions of the time-independent Schr\"odinger equation.  
In particular, it was shown that an intrinsic vector potential arises whenever the total electronic state carries a non-vanishing current [as per Eq.~\eqref{eq:Arelj} below].

For the $m$-th eigenstate of the Schr\"odinger equation \eqref{eq:TISE}, the electronic vector potential is defined as
\begin{equation}\label{eq:Adef}
\mathbf{A}_m(\rr) = - i\,\langle \Phi_m | \nabla_\rr \Phi_m \rangle_{\ssm\rr}
\end{equation}
where the exact factorization of the electronic wavefunction [Eq.\eqref{eq:EFWF}] is generalized to excited states.
Equation \eqref{eq:Adef} is in line with the definition of the time-dependent vector potential reported in the literature \cite{Gross_PRL2017, Schild_PRR2020, kocak2022eef}; however, writing the conditional amplitude in polar form,
\begin{equation}
\Phi_m(\sigma,2,\dots,N;\rr)
= e^{i\theta_m(\sigma,2,\dots,N;\rr)}R_m(\sigma,2,\dots,N;\rr)
\end{equation}
one finds
\begin{equation}
\mathbf{A}_m(\rr)
= \langle R_m | \nabla_\rr \theta_m | R_m \rangle_{\ssm\rr}
\end{equation}
which makes explicit that the electronic vector potential is entirely determined by the phase $\theta_m$ of the conditional amplitude.
Furthermore, this vector potential is directly related to the paramagnetic current density of the many-electron wavefunction via~\cite{GiaGorAgo-CPC-24} 
\begin{equation}\label{eq:Arelj}
\mathbf{A}_m(\rr)
= \frac{\mathbf{j}_p(\rr)}{\rho(\rr)}
\end{equation}
where $\mathbf{j}_p(\rr)=N\,\mathrm{Re}\langle\Psi|-i\nabla_\rr\Psi\rangle{\ssm\rr}$.

\begin{figure}
\centering
\includegraphics[width=0.81\linewidth]{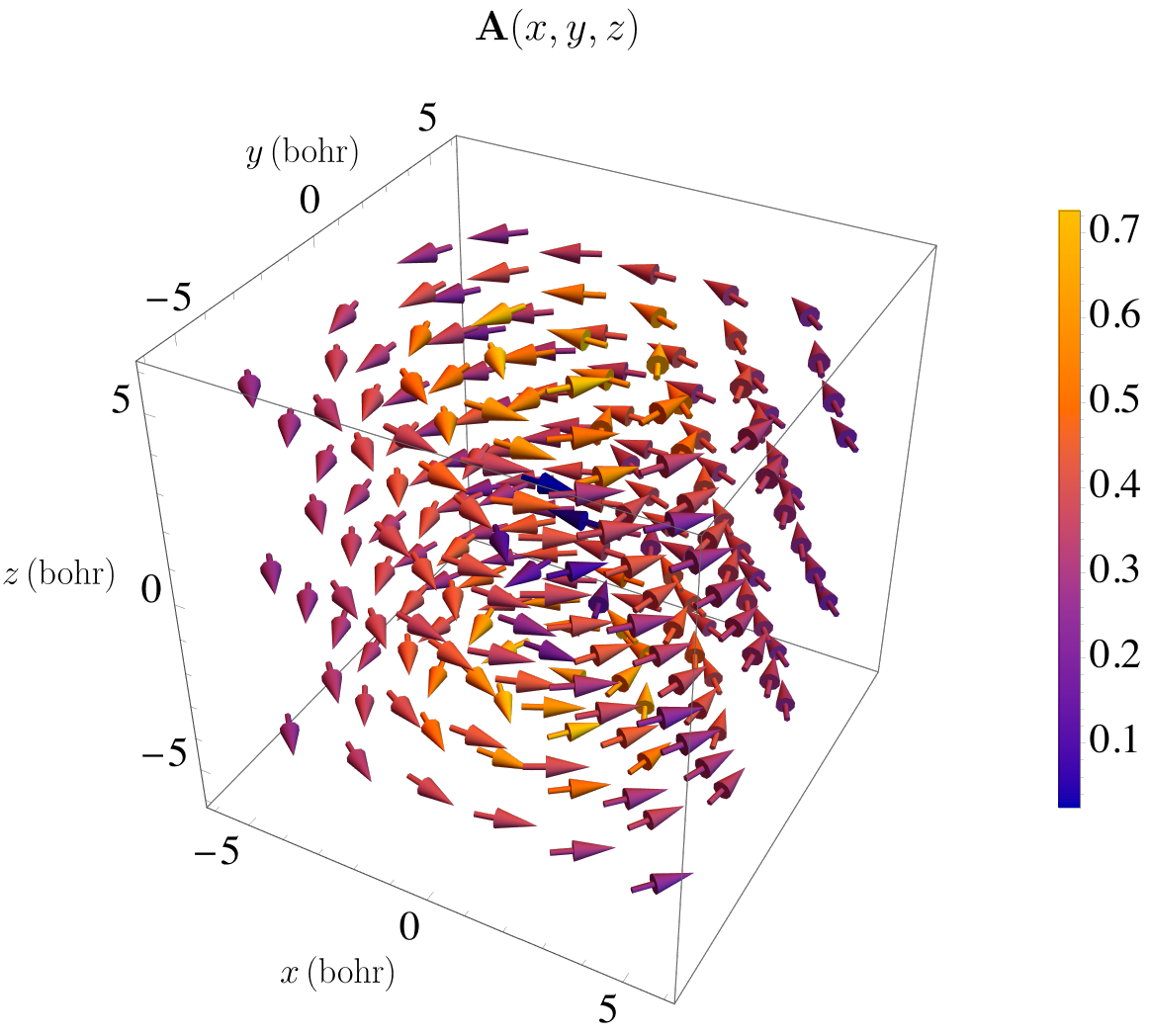}
	\caption{\small{Electronic vector potential 
    for two non-interacting fermions in a triplet state in Cartesian coordinates $x, \, y, \, z$ for nuclear charge $Z=1$. 
Color coding reflects the magnitude of the vector fields at each point. Adapted from Giarrusso, S., Gori-Giorgi, P., \& Agostini, F. (2024). ChemPhysChem, 25(18), e202400127, licensed under CC BY.}}
	\label{fig:Avectorplot}
 \end{figure}

To explicitly demonstrate the physical relevance of the electronic vector potential in at least one case, the triplet state of two non-interacting fermions in a Coulomb potential was analysed.
In this case, the associated vector potential (shown in Fig.~\ref{fig:Avectorplot}) has a non-vanishing curl \cite{GiaGorAgo-CPC-24}, and therefore cannot be removed by a gauge transformation, proving that it is not an artifact of gauge choice.
Note that the concept of a time-independent vector potential stemming from the EEF also appears in Ref.~\cite{TseVay-AP-12}, however in this work only the irrotational case is discussed.

\section{Exact photon-electron-nuclear factorization}\label{sec:cavities}
\begin{figure}
    \centering
\includegraphics[width=\columnwidth]{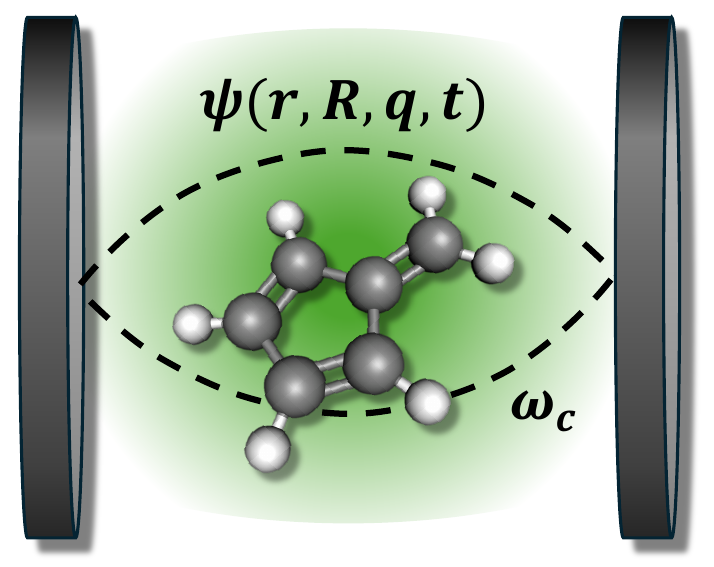}
    \caption{Pictorial representation of a molecule in an optical cavity with frequency $\omega_c$. The photon-electron-nuclear system is described by the many-body wavefunciton $\Psi(\elr,\nucR,\phq,t)$.}
    \label{fig:cavity}
\end{figure}
Molecules within an optical cavity are able to coherently exchange energy with quantized electromagnetic modes (Fig.~\ref{fig:cavity}). This framework, known as molecular cavity quantum electrodynamics (cQED), allows one to describe a rich variety of phenomena. For example, in the weak-coupling regime, where light-matter interactions are smaller than the system's loss rates, the cavity can suppress or enhance spontaneous emission through the Purcell effect~\cite{Purcell1946}. By contrast, under strong coupling, where the interaction dominates over losses, the molecular excitations and the cavity photons hybridize into the so-called \textsl{polaritons}. These hybrid light-matter states reshape the molecular potential energy surfaces, allowing one to alter intrinsic properties such as photochemical reactivity, giving rise to the field often termed \textit{polaritonic photochemistry}~\cite{Hutchison2012, Ebbesen_ACR2016, Rashidi2025, Garg2025, Fregoni2018, Groenhof2024, Groenhof2022, Feist2018, Feist2022, Subotnik2022-2}. 

Since the first realizations of the effect of strong light-matter coupling on the photochemical reactivity, many theories and computational methods for electronic structure~\cite{Angelico2023, Riso2022, Haugland2020, Schfer2022, Wang2019, Ruggenthaler2014, Flick2017_2, Wickramasinghe2025, Bonini2022, Bonini2024} and for molecular dynamics~\cite{De2024,Krupp2025, Hoffmann2019, Hoffmann2019_jun, Maitra_JCP2020, Hu2023, Sokolovskii2024, Hu2025,Tichauer2021,Vendrell2018,Li2021,Li2021_2,Rana2023,Yu2022,Li2022} have been generalized to the field of cQED~\cite{Kowalewski2016,Galego2015,Galego2016,Galego2017,Fischer2023}. Extensive reviews on cQED as well as its applications can be found in Refs.~\cite{Mandal2023,Flick2017_1,Ribeiro2018,Yuen-Zhou2026-tt, Ruggenthaler2023,Taylor2025, Rubio2018, Mukamel2023}. 

\subsection{Exact factorization in the presence of photons}
The molecular cQED Hamiltonian $\hat H(\elr,\nucR,\phq)$ describing a system of interacting photons, electrons and nuclei is
\begin{align}
\label{eq:fullcavH}
\hat H =\hat T_n+\hat H_{el} (\elr,\nucR)+ \hat T_p+\hat V_p (\phq) + \hat H_{n,el,p} (\elr,\nucR,\phq)
\end{align}
This Hamiltonian is in the literature also referred to as the Pauli-Fierz non-relativistic QED Hamiltonian in the dipole gauge~\cite{Mandal2023,Woolley2020,Taylor2025} and is rigorously derived from the minimal coupling Hamiltonian in the Coulomb gauge by applying the Power-Zienau-Woolley Gauge transformation~\cite{Power1959} and a unitary phase transformation under the assumption that the dimension of the molecular system is much smaller than the length of the cavity, \ie the long wavelength approximation. This derivation is discussed in detail in, for instance, Ref.~\cite{Mandal2020}.

In Eq.~(\ref{eq:fullcavH}), the matter contribution is $\hat H_M (\elr,\nucR)=\hat T_n+\hat H_{el} (\elr,\nucR)$, with the nuclear kinetic energy operator $\hat T_n$ and the electronic Hamiltonian $\hat H_{el}$ containing the electronic kinetic energy and all the interactions. Quantization of the electromagnetic field yields the photon Hamiltonian $\hat H_p(\phq)=\hat T_p+\hat V_p (\phq)$ essentially represented in terms of harmonic oscillators $\hat H_p(\phq)=\frac{1}{2}(\sum^{N_p}_\alpha \hat p ^2_\alpha+\omega_\alpha^2\hat q^2_\alpha)$ with unitary masses and frequencies $\omega_\alpha$. The sum over $\alpha$ runs over both possible transverse polarizations of the electromagnetic field, \ie $N_p$. Furthermore, the photons' degrees of freedom are represented by the photonic displacement operator $\hat q_\alpha=\sqrt{\hbar/(2\omega_\alpha)}(\hat a_\alpha^\dagger+\hat a_\alpha)$ and by its conjugated momentum operator $\hat p_\alpha=i\sqrt{\hbar/(2\omega_\alpha)}(\hat a^\dagger_\alpha-\hat a_\alpha)$, which are related to the electric field and to the magnetic field, respectively. The coupling Hamiltonian between the photons and the electron-nuclear system, \ie the matter, is $\hat H_{pM}=\sum^{N_p}_\alpha \omega_\alpha \lambda_\alpha \hat q_\alpha(e\sum^{N_n}_\nu Z_\nu \hat\nucR_\nu -e\sum^{N_e}_i \hat\elr_i)$, with $\lambda_\alpha$ the coupling strength between light and matter. The indices $\nu$ and $i$ run over the nuclei (with charges $Z_\nu e$) and the electrons. Note that, in writing Eq.~(\ref{eq:fullcavH}) we neglected the so-called self-polarization term, depending only on electron-nuclear degrees of freedom, which is usually required to ensure translational invariance in the field-free case and to ensure that $\hat H$ is bound form below~\cite{Rokaj2018}. However, for systems with few photon modes the contribution of this term was observed to be small and can be neglected in same cases~\cite{Maitra_JCP2020,Tichauer2021,Agostini_JCP2024_2}.

The time-dependent Schr\"odinger equation with Hamiltonian~(\ref{eq:fullcavH})
\begin{align}
i\hbar \partial_t\Psi(\elr,\nucR,\phq,t) =\hat H(\elr,\nucR,\phq)\Psi(\elr,\nucR,\phq,t)
\end{align}
governs the evolution the photon-electron-nuclear (PEN) wavefunction, which can be, following the strategy of the exact factorization, factored as the product of a marginal amplitude and of a conditional amplitude.

Various possibilities for the separation into marginal and conditional variables have been proposed~\cite{Maitra_EPJB2018, Tokatly_EPJB2018, Maitra_JCP2020, Maitra_JCP2021, Maitra_JCP2022, Agostini_JCP2024_2, Maitra_PRL2019}, which can be related to other theoretical frameworks that have been developed for cQED, even beyond the exact factorization. Specifically, the literature reports three different perspectives:
\begin{enumerate}
\item the polaritonic Born-Oppenheimer (pBO) approach, which can be related to the choice of associating the nuclear positions to the marginal wavefunction~\cite{Schfer2018,De2024,Fregoni2020,Hu2023,Sokolovskii2024,Hu2025}
\item the photon Born-Oppenheimer (qBO) approach, which corresponds to associating the photonic degrees of freedom to the marginal wavefunction~\cite{Maitra_JCP2020}
\item the cavity Born-Oppenheimer (cBO), which is related to considering the photon-nuclear degrees of freedom as marginal variables ~\cite{Fischer2023, Flick2017_1, Flick2017_2, Kowalewski2016, Wickramasinghe2025, Bonini2022, Bonini2024,Angelico2023}
\end{enumerate}
The above mentioned approaches to decomposed the PEN wavefunction usually adopt the Born-Oppenheimer (Born-Huang) representation of the wavefunction, meaning that $\Psi(\cox,\maX,t)=\sum_n \chi_n(\maX,t)\varphi_n(\cox;\maX)$, employing the generalized formalism introduced in Sec.~\ref{sec: theory}. The basis formed of conditional-like amplitudes, \ie $\varphi_n(\cox;\maX)$, is chosen such that the elements are the eigenstates of the Hamiltonian operators defined as $\hat H-\hat T_n$ in pBO, of $\hat H-\hat T_p$ in qBO, and of $\hat H-\hat T_n-\hat T_p$ in cBO. 

Employing the concepts of pBO, qBO and cBO referring to the general theory of photochemistry in the strong light-matter coupling regime, also for the exact photon-electron-nuclear factorization (EPENF) we can introduce the concepts of 
\begin{enumerate}
    \item polaritonic exact factorization (pEF)~\cite{Maitra_PRL2019,Maitra_JCP2021}
    \item photon exact factorization (qEF)~\cite{Maitra_EPJB2018, Maitra_JCP2022}
    \item cavity exact factorization (cEF)~\cite{Agostini_JCP2024_2}
\end{enumerate}
and indeed, these approaches have been developed and applied in the EPENF literature.

The pEF approach of Refs.~\cite{Maitra_PRL2019,Maitra_JCP2021} is reviewed in Sec.~\ref{sec: pEF} and refers to the treatment of the nuclear degrees of freedom as the marginal variables, with the photonic and electronic degrees of freedom appearing in the conditional amplitude. Such a perspective would allow one to straightforwardly extend nonadiabatic molecular dynamics schemes to the study of polaritonic photochemistry, focusing on the description of ``standard'' nuclear dynamics under the effect of hybrid photon-electronic potentials, even though these route has not yet been fully investigated within EPENF.

In qEF, instead, as done in Refs.~\cite{Maitra_EPJB2018, Maitra_JCP2022}, the photonic degrees of freedom are associated to the marginal amplitude, with the electronic and nuclear variables appearing in the conditional amplitude. We will present this perspective on EPENF in Sec.~\ref{sec: qEF}.

The cEF approach employed in Ref.~\cite{Agostini_JCP2024_2} and presented in Sec.~\ref{sec: cEF} treats as marginal variables both photonic and nuclear degrees of freedom, coupled to the conditional electronic coordinates, and revealed potential for cost effective simulations, in particular when adapting nonadiabatic molecular dynamics schemes to the treatment of the photonic degrees of freedom in terms of trajectories.

Finally, let us mention the time-independent approach that was presented in Ref.~\cite{Tokatly_EPJB2018}, which we refer to as \textsl{cavity electron exact factorization} (ceEF), and which treats as marginal variables the electronic degrees of freedom, similarly to Ref.~\cite{Suzuki_PRA2014}. This approach has been applied in the clamped-nuclei approximation, thus the nuclear degrees of freedom do not appear explicitly in the problem, and focused on determining the electronic eigenfunction in the presence of strong light-matter coupling, as described in Sec.~\ref{sec: ceEF}.

Overall, we have identified four distinct approaches in the literature based on the exact factorization which have been used to study the PEN problem with a cQED framework. The approaches are summarized in Table~\ref{tab:typesofCavityEF} -- and compared to the original exact electron-nuclear factorization (EF) in the first row of the table -- and will be discussed in detail below. The overarching goal of introducing these perspectives in cQED is to achieve some fundamental understanding of the most efficient and accurate procedure to simulate the electron-nuclear dynamics in nonadiabatic processes under the effect of strong coupling with light, possibly exploiting the knowledge acquired over the years with the exact factorization on related problems but in cavity-free conditions. The next sections illustrate the work done and ongoing in this field.

\begin{table}
    \centering
    \caption{Summary table of the studies employing the exact factorization in combination with cQED. 
    \textit{c.n.} refers to the clamped nuclei approximation.
    }
         \label{tab:typesofCavityEF} 
    \begin{tabular}{lcccc}
        \toprule
        Type & Marginal& Conditional & Refs. &  Section\\
        \midrule
                EF & $\chi(\nucR,t)$  & $\Phi (\elr,t;\nucR)$ & \cite{Gross_PRL2010, Gross_JCP2012, Gross_PTRSA2014} \\
                \midrule
        pEF & $\chi(\nucR,t)$& $\Phi (\phq,\elr,t;\nucR)$ & \cite{Maitra_PRL2019,Maitra_JCP2021} & \ref{sec: pEF} \\
         qEF & $\chi(\phq,t)$& $\Phi (\nucR,\elr,t;\phq)$ & \cite{Maitra_JCP2022,Maitra_EPJB2018} & \ref{sec: qEF}  \\
        cEF & $\chi(\nucR,\phq,t)$&  $\Phi(\elr,t; \nucR,\phq)$ & \cite{Agostini_JCP2024_2} & \ref{sec: cEF} \\
        ceEF (\textit{c.n.}) & $\chi(\elr,t)$& $\Phi (\phq,t;\elr)$ &  \cite{Tokatly_EPJB2018} & \ref{sec: ceEF} \\
        \bottomrule
    \end{tabular}
\end{table}

\subsection{Polaritonic exact factorization}\label{sec: pEF}
The pEF is a very intuitive generalization of the exact electron-nuclear factorization since it allows to visualize the PEN dynamics in terms of the marginal nuclear wavefunction whose evolution is governed by the time-dependent scalar and vector potentials representing the effect of the photon-dressed electrons, using the expression in the second row of Table~\ref{tab:typesofCavityEF} for the PEN wavefunction
\begin{align}
    \Psi(\elr,\nucR,\phq,t)=\chi(\nucR,t)\Phi (\phq,\elr,t;\nucR)
\end{align}
Therefore, the key quantities in pEF are the polaritonic time-dependent vector potential (pTDVP) defined as
\begin{align}\label{eq: tdvp in pEF}
\TDVP^{\mathrm{pEF}}_\nu(\nucR,t)=\braket{\Phi(t;\nucR)|-i\hbar\nabla_\nu\Phi(t;\nucR)}_{\elr,\phq}    
\end{align}
and the polaritonic time-dependent potential energy surface (pTDPES) 
\begin{align}\label{eq: tdpes in pEF}
\TDPES^{\mathrm{pEF}}(\nucR,t)=\TDPES_{c}^{\mathrm{pEF}}(\nucR,t)+\TDPES^{\mathrm{pEF}}_{geo}(\nucR,t)+\TDPES^{\mathrm{pEF}}_{GD}(\nucR,t)    
\end{align}
whose components are defined as
\begin{align}
    \TDPES_{c}^{\mathrm{pEF}}(\nucR,t)&=\braket{\Phi|\hat H - \hat T_n|\Phi}_{\elr,\phq}\label{eq: bo tdpes in pEF}\\
    \TDPES^{\mathrm{pEF}}_{geo}(\nucR,t)&= \sum_\nu \frac{\hbar^2\langle \nabla_\nu \Phi|\nabla_\nu \Phi\rangle_{\elr,\phq}}{2M_\nu}-\frac{[\TDVP^{\mathrm{pEF}}_\nu(\nucR,t)]^2}{2M_\nu}\label{eq: geo tdpes in pEF}\\
    \TDPES^{\mathrm{pEF}}_{GD}(\nucR,t)&=\braket{\Phi|-i\hbar \partial_t|\Phi}_{\elr,\phq}\label{eq: gd tdpes in pEF}
\end{align}
in strict analogy to the electron-nuclear case. Note that Eqs.~(\ref{eq: bo tdpes in pEF}) and~(\ref{eq: geo tdpes in pEF}) are gauge invariant, whereas Eq.~(\ref{eq: gd tdpes in pEF}) depends on the choice of gauge, as well as Eq.~(\ref{eq: tdvp in pEF}).

Beyond EPENF, the pBO picture has been employed extensively in the literature~\cite{Schfer2018,De2024,Fregoni2020,Hu2023,Sokolovskii2024,Hu2025} as it allows one to formulate the PEN dynamics adopting the concepts of polaritonic potential energy surfaces (PoPES), defined as the eigenvalues of the stationary Schr\"odinger equation
\begin{align}
    \left(\hat H(\elr,\nucR,\phq) - \hat T_n\right)\varphi_n(\phq,\elr;\nucR) = E_n^{\mathrm{pBO}}(\nucR)\varphi_n(\phq,\elr;\nucR)
\end{align}
and introducing a Born-Huang-like representation of the PEN wavefunction, namely $\Psi(\elr,\nucR,\phq,t)=\sum_n\chi_n(\nucR,t)\varphi_n(\phq,\elr;\nucR)$. 

\begin{figure}
    \centering
\includegraphics[width=\columnwidth]{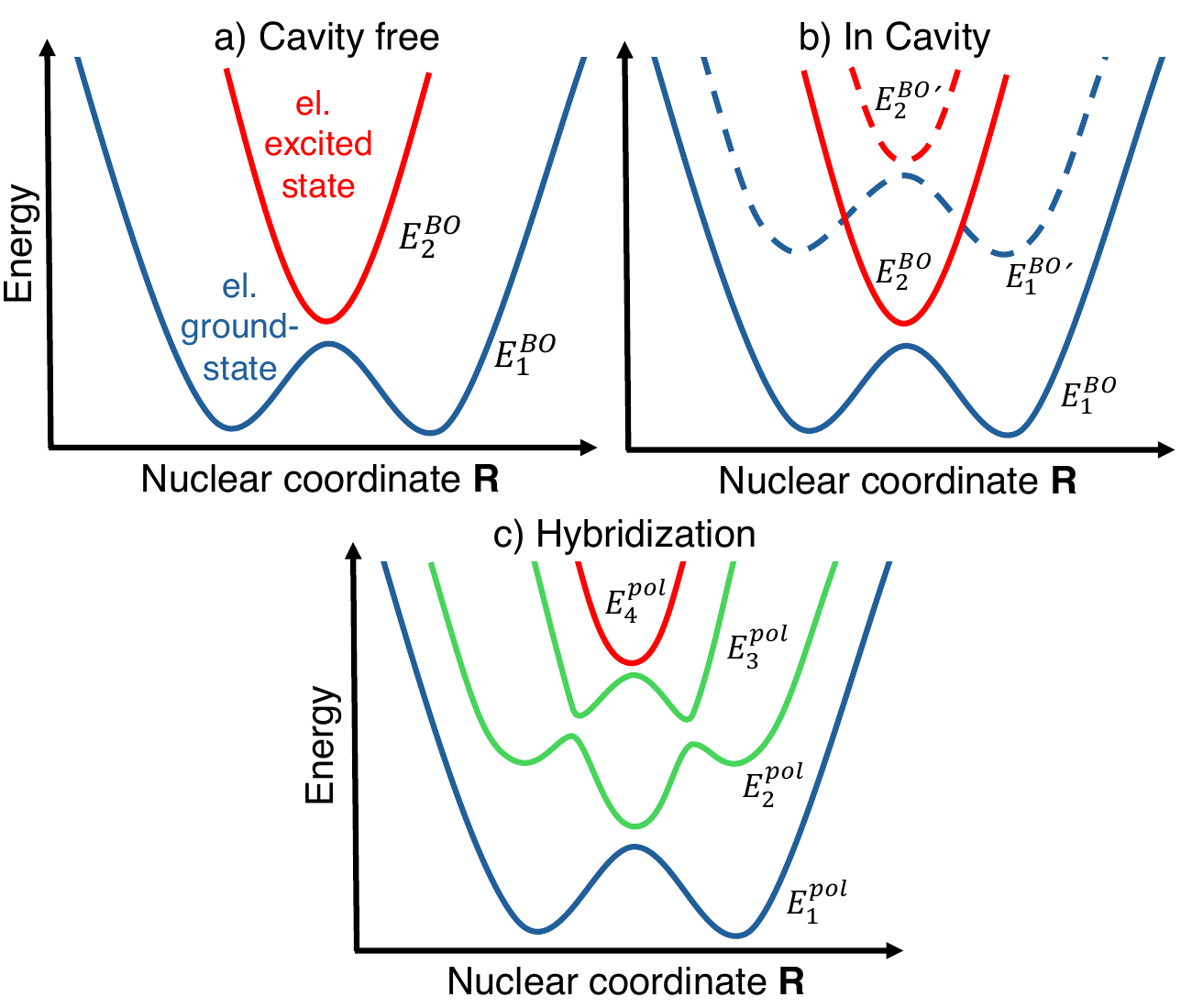}
    \caption{Schematic representation of the cavity-free electronic BOPES (panel a), indicated as $E^ {BO}_1$ and $E^{BO}_2$, of the new energy states appearing inside the cavity due to the coupling between the molecular excitation and light (panel b), indiacted as $E^ {BO\prime}_1$ and $E^ {BO\prime}_2$, of the hybrid PopES (panel c), indicated as $E_n^{pol}$ with $n=1,\ldots,4$.}
    \label{fig:polaritonicsurf}
\end{figure}

Figure~\ref{fig:polaritonicsurf} schematically depicts the changes of the potential energy landscape from the cavity-free molecule to the strong light-matter coupling regime, the latter accompanied by the emergence of the PoPESs (neglecting self-polarization terms). Outside the cavity, the molecule is characterized by the Born-Oppenheimer (BO) potential energy surfaces (PESs), for the electronic ground (blue) and excited state (red), which are functions of the nuclear coordinate $\nucR$ (Fig.~\ref{fig:polaritonicsurf}a). When the molecule finds itself in the presence of photonic modes inside the cavity, ``additional'' states appear, since the photon-electronic Hilbert space is the vector product of the Hilbert spaces for the electrons and for the photons (using, for instance, the discrete number representation). This is shown in Fig.~\ref{fig:polaritonicsurf}b, where we observe an increase of the number of PESs with respect to Fig.~\ref{fig:polaritonicsurf}a. Each electronic PES of Fig.~\ref{fig:polaritonicsurf}a is replicated into a ladder of photon-dressed states, separated by the cavity frequency (dashed curves in Fig.~\ref{fig:polaritonicsurf}b, where we only considered the zero-photon and the one-photon states). In the presence of coupling between the photonic modes and the electronic molecular excitations, hybrid photon-electronic states, \ie the polaritonic states, are formed. Hybridization occurs in the regions of crossings of PES, yielding new PoPESs (green lines in Fig.~\ref{fig:polaritonicsurf}c), which qualitatively reshape the potential energy landscape and, thus, profoundly affect the molecular photodynamics.

\begin{figure}
    \centering
\includegraphics[width=\columnwidth]{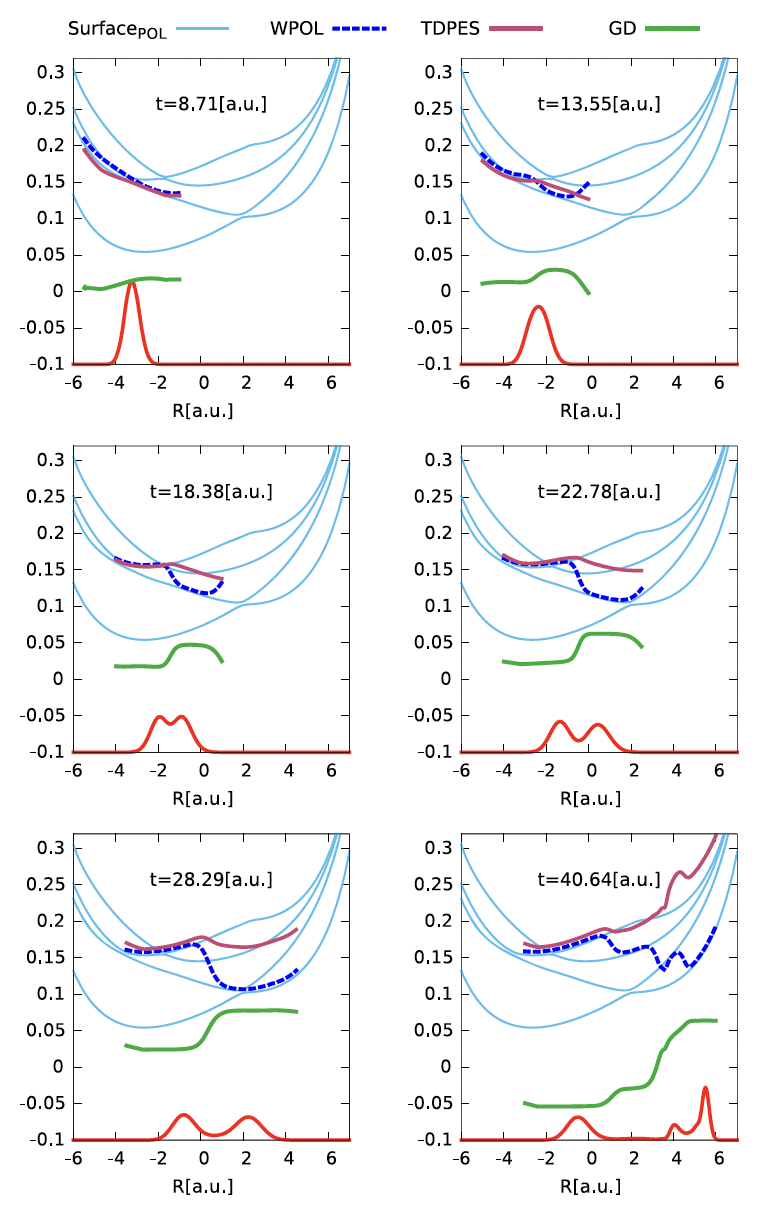}
    \caption{Nuclear dynamics of the Shin-Metiu model of Ref.~\cite{Maitra_PRL2019} in the regime of strong coupling with light, with snapshots of the marginal nuclear density (red curves) at different time steps indicated in the panels. The PoPESs (Surface$_{\mathrm{POL}}$ in the figure, light blue curves) do not depend on time and are used as reference for the pTDPES (brown curves), for its gauge-invariant contribution $\TDPES^{\mathrm{pEF}}_c(\nucR,t)$ (WPOL in the figure, dashed dark blue curves) and for its gauge-dependent part $\TDPES_{GD}^{\mathrm{pEF}}(\nucR,t)$ (GD in the figure, green curves). Reproduced from Lacombe, L., Hoffmann, N. M. \& Maitra, N. T. (2019). Physical Review Letters, 123(8), 083201 with the permission of APS.}
    \label{fig: shin-metiu in cavity}
\end{figure}

The pEF with the support of the PoPESs have been employed in some model studies on a proton coupled electron transfer process, using for instance the Shin-Metiu model~\cite{Shin1995,Shin1996}. This model features a proton moving in one dimension between two fixed positive charges (Coulomb interaction) and an electron moving as well in one dimension interacting with all positive charges via soft Coulomb potentials. The simulated dynamics, \eg in Ref.~\cite{Maitra_PRL2019}, starts with the proton on the left side close to the left fixed charge with the electron in the first excited state. As also presented in Ref.~\cite{Gross_PRL2013}, this electronic state is a charge-transfer state, since the electron is localized on the right fixed charge. In the course of the dynamics, the proton moves to the right but due to nonadiabatic effects, the electron partially transfers to the left fixed charge, since both the ground state and the first excited state become populated. At the end of the dynamics, the proton fully transfers to the right. In Ref.~\cite{Maitra_PRL2019}, strong coupling of this system with a cavity mode alters this dynamics by partially preventing the proton transfer, an effect that can be clearly observed via the analysis of the pTDPES. The analysis of the nuclear dynamics in a series of papers by Maitra and co-workers~\cite{Maitra_PRL2019, Maitra_JCP2021, Maitra_JCP2020} especially aims to investigate the properties of the pTDPES in the course of the nuclear dynamics. Figure~\ref{fig: shin-metiu in cavity} of Ref.~\cite{Maitra_PRL2019} shows the results of such dynamics representing a cavity-induced suppression of proton coupled electron transfer (with $N_n=1$, $N_e=1$, $N_p=1$). In this work, since the marginal amplitude is chosen to be the nuclear wavefunction, the mass $M_\nu$ of Eq.~(\ref{eq: geo tdpes in pEF}) is the proton mass. Following the arguments validated in Ref.~\cite{AgostiniEich_JCP2016}, $\TDPES^{\mathrm{pEF}}_{geo}$ can be neglected in comparison to the other two contributions in the pTDPES, namely Eqs.~(\ref{eq: bo tdpes in pEF}) and~(\ref{eq: gd tdpes in pEF}). The gauge is chosen such that $\TDVP^{\mathrm{pEF}}_\nu=0$ so that the nuclear dynamics is fully encoded in $\TDPES^{\mathrm{pEF}}\approx \TDPES^{\mathrm{pEF}}_{c}+\TDPES^{\mathrm{pEF}}_{GD}$. Note that, as shown in Ref.~\cite{IBELE2024188}, this a possible and general choice of gauge when working in a one-dimensional nuclear space. 

In Fig.~\ref{fig: shin-metiu in cavity}, the PoPES (light blue lines) are compared to the pTDPES (dark violet line), to the gauge dependent part $\TDPES_{GD}^{\mathrm{pEF}}$ (green line, GD in the figure) and to the \textsl{polaritonic term}, \ie $\TDPES_{c}^{\mathrm{pEF}}$ according to our definitions (dashed dark blue line,  WPOL in the figure). The nuclear density (red line) starts as a Gaussian centered at $R=-2$~a.u. (Franck-Condon region) in the second electronic states, \ie the first excited state. Note that, in that region, the PoPES is identical to a pure electronic potential and the system is initialized in a zero-photon state. As the evolution starts, the nuclear density moves to the right following the slope of the polaritonic surfaces. Around $t = 18.38$~a.u., a step in the pTDPES begins to develop, forcing the density to split and, thus, partially to be reflected back towards the Franck-Condon region. This reflected part remains, then, localized in a potential well formed on the left-hand side ($t=22.78$~a.u. and later times), preventing the complete transfer of nuclear density to the right. The other part of the wavepacket, \ie the transmitted part, instead, was found to behave similar to the cavity-free dynamics (as analyzed as well in Ref.~\cite{Maitra_PRL2019}), propagating to the right-hand side ($t=40.64$~a.u.), meaning that it moves towards the right branching into different electronic states. Further analysis of the dynamics \cite{Maitra_PRL2019,Maitra_JCP2021} shows that comparison between the $\TDPES_{c}^{\mathrm{pEF}}$ and the PoPESs yields information about the occupations of the polaritonic states, in terms of space-and-state resolved nuclear density. In addition, the contribution from $\TDPES_{GD}^{\mathrm{pEF}}$ adjusts the energy in a stepwise manner to yield the correct nuclear dynamics. From this in-depth analysis employing a fully dynamical perspective encoded in the pTDPES, it is clear that while polaritonic surfaces provide a general understanding of the energy landscape, they do not reflect and explain the complex nuclear dynamics. 

\subsection{Photon exact factorization}\label{sec: qEF}
One of the earlier studies to apply the exact factorization in the framework of cQED was presented by Hoffman \textit{et al.} in Ref.~\cite{Maitra_EPJB2018}, focusing on the (marginal) dynamics of the photonic degrees of freedom. The motivation to adopt such a perspective, also motivated by some earlier studies~\cite{Hoffmann2019}, was to employ a cost-effective trajectory-based description of the evolution of the photons' displacements and conjugated momenta, since the photon Hamiltonian $\hat H_p$ is ``simply'' a sum of uncoupled harmonic oscillators. The hypothesis driving this study was that classical trajectories from an initial Wigner-sampled Gaussian probability distribution yield an exact harmonic quantum dynamics~\cite{Heller1976}. Also, within the dipole approximation, the linear light-matter coupling is preserved. 

The qEF approach was introduced, then, in Ref.~\cite{Maitra_EPJB2018} employing the expression for the PEN wavefunction in the third row of Table~\ref{tab:typesofCavityEF}
\begin{align}
\Psi(\elr,\nucR,\phq,t)=\chi(\phq,t)\Phi(\nucR,\elr,t;\phq)
\end{align}
It is wroth noting that, despite the fact that the photon Hamiltonian is purely harmonic, the coupling to the matter, \ie electrons and nuclei, translates into a qTDPES which is strongly anharmonic. Such qTDPES is the potential that, together with the qTDVP, drives the evolution of the photonic degrees of freedom. As in Eq.~\eqref{eq: tdpes GI+GD}, the qTDPES reads 
\begin{align}\label{eq: tdpes in qEF}
\TDPES^{\mathrm{qEF}}(\phq,t)=\TDPES_{c}^{\mathrm{qEF}}(\phq,t)+\TDPES^{\mathrm{qEF}}_{geo}(\phq,t)+\TDPES^{\mathrm{qEF}}_{GD}(\phq,t)
\end{align}
and, similarly to Eq.~(\ref{eq: tdpes in pEF}), it is the sum of three contributions, namely
\begin{align}
    \TDPES_{c}^{\mathrm{qEF}}(\phq,t)&=\braket{\Phi|\hat H - \hat T_p|\Phi}_{\elr,\nucR}\label{eq: bo tdpes in qEF}\\
    \TDPES^{\mathrm{qEF}}_{geo}(\nucR,t)&= \sum_\alpha \frac{\hbar^2}{2}\langle \nabla_\alpha \Phi|\nabla_\alpha \Phi\rangle_{\elr,\nucR}-\frac{[\TDVP^{\mathrm{qEF}}_\alpha(\phq,t)]^2}{2}\label{eq: geo tdpes in qEF}\\
    \TDPES^{\mathrm{qEF}}_{GD}(\phq,t)&=\braket{\Phi|-i\hbar \partial_t|\Phi}_{\elr,\nucR}\label{eq: gd tdpes in qEF}
\end{align}
that are formally very similar to Eqs.~(\ref{eq: bo tdpes in pEF}) to~(\ref{eq: gd tdpes in pEF}). Note that in Eq.~(\ref{eq: geo tdpes in qEF}) the sum runs over the field polarizations and the masses are the effective unitary masses of the harmonic oscillators representing the photons. In addition, the pTDVP analogous to Eq.~(\ref{eq: tdvp in pEF}) is
\begin{align}\label{eq: tdvp in qEF}
    \TDVP^{\mathrm{qEF}}_\alpha(\phq,t)=\braket{\Phi(t;\phq)|-i\nabla_\alpha\Phi(t;\phq)}_{\elr,\nucR}
\end{align}
The strong anharmonicity of the qTDPES mainly arises due to the presence of $\TDPES_{geo}^{\mathrm{qEF}}(\phq,t)$, as it was also argued later on in the framework of cEF in Ref.~\cite{Agostini_JCP2024_2}. By contrast to pEF, this term cannot be neglected in the expression of the qTDPES because the arguments based on the small electron-nuclear and photon-nuclear mass ratios~\cite{AgostiniEich_JCP2016} cannot be invoked. 

The numerical applications reported in Ref.~\cite{Maitra_EPJB2018} focused on a two-electronic-state model Hamiltonian, where the electronic states are separated by an energy gap $E$. The model can be viewed as $\phq$-resolved realization of the quantum Rabi model~\cite{Rabi1936, Rabi1937}, where the two states are coupled via a transition dipole moment $D$ with a coupling strength $\lambda$. It is worth noting, in addition, that the nuclear degrees of freedom do not appear explicitly in the dynamics, thus we can consider this case as a particular realization of qEF with clamped nuclei.

\begin{figure*}
    \centering
\includegraphics[width=\textwidth,page=1,trim={0cm 2cm 0cm 0cm},clip]{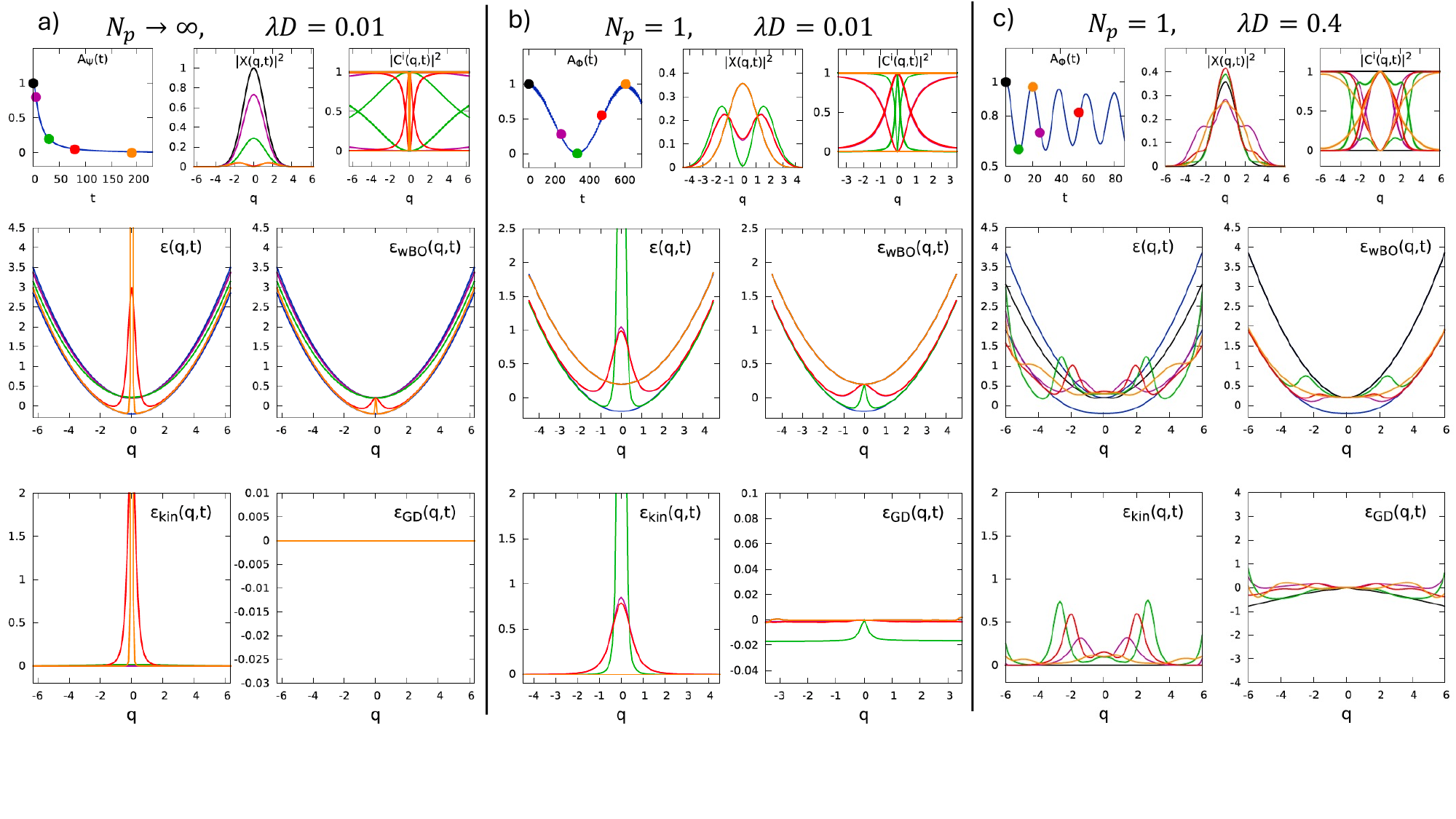}
    \caption{Photon dynamics in a two-level system coupled to a cavity. Top panels: Time-dependent autocorrelation function $A_\Psi(t)|$ from which various time steps are selected as colored dots, at which the following figures are shown (left); Marginal photon density at the time steps identified in $A_\Psi$ (center); Electronic coefficients from the Born-Huang-like expansion of the electronic conditional amplitude from Eq.~\eqref{eq.expcofqBO} (right). 
    Middle and bottom panels: one-dimensional qTDPES ($\varepsilon(q,t)$) along with its contributions, namely $\epsilon_{wBO}$, $\epsilon_{kin}$ and $\epsilon_{GD}$, which correspond to Eqs.~\eqref{eq: bo tdpes in qEF},~\eqref{eq: geo tdpes in qEF} and~\eqref{eq: gd tdpes in qEF}, respectively. The qTDPES is shown at the time steps identified by the colored dots in $A_\Psi$. Panel a): Dynamics in the Wigner-Weisskopf limit for weak coupling $\lambda D=0.01$. Panel b): Same as a) but for $N_p=1$. Panel c): Same as b) but for strong coupling $\lambda D=0.4$. 
    Adapted from Hoffmann,  N. M., Appel, H., Rubio, A., \& Maitra, N. T. (2018). The European Physical Journal B, 91(8), 180, licensed under CC BY.}
    \label{fig:cavityhoff}
\end{figure*}

In the Wigner-Weisskopf limit, \ie infinite number of modes $N_p\to \infty$, the exact result of this model is known~\cite{Scully1997-ba}. From the solution, the qTPDES can be determined. However, since it is infinitely dimensional, insightful observations can only be drawn by analyzing cuts of the infinite-dimensional qTDPES along some representative modes, as for instance the resonant mode $q_r$  which is shown in Fig.~\ref{fig:cavityhoff}a. In the case of a single resonant mode $N_p=1$ (with $\omega_c=E$), instead, shown in Fig.~\ref{fig:cavityhoff}b and ~\ref{fig:cavityhoff}c, the analysis of the one-dimensional qTDPES becomes straightforward. 

The dynamics studied in Ref.~\cite{Maitra_EPJB2018}, as well as in Ref.~\cite{Maitra_JCP2022} later on, employing the qEF exemplifies the coupled photon-electron dynamics with clamped nuclei. The electronic system is initialized in the excited state, with the photon density chosen as a Gaussian centered at the origin: the photon field is associated to the zero-photon state. In the absence of coupling between the electronic states via the transition dipole moment, the system remains in this eigenstate. However, the strong coupling of the photon mode in resonance with the electronic energy gap yields an transition to the ground state accompanied by the emission of a photon, which is then reabsorbed giving rise to typical Rabi oscillations. The emission of the photon can be observed in the photon density which develops a bimodal distribution. In Ref.~\cite{Maitra_EPJB2018} the ultrastrong coupling regime as well as the strong coupling regime in the presence of an infinite number of modes are also investigated and connection are drawn among the different dynamics.

Figure~\ref{fig:cavityhoff} reports the numerical results of Ref.~\cite{Maitra_EPJB2018} and shows various qEF quantities along the dynamics. We recall that, in this study, the qEF is applied in the clamped nuclei approximation, thus the problem essentially describes two electronic states coupled to $N_p$ cavity modes. The three panels, a, b and c of Fig.~\ref{fig:cavityhoff} show various time-dependent quantities: the autocorrelation function $A_\Psi(t)=|\int d\phq\,\Psi^*(\elr,\phq,t=0)\Psi(\elr,\phq,t))|$, where some time steps are identified by colored dots such that various time-dependent $\phq$-dependent quantities are shown in the other panels (top left); marginal photon density as function of $\phq$ at the times indicated by the dots $|\chi(\phq,t)|^2$ (top center); $\phq,t$-dependent electronic coefficients at various time steps (top right); qTDPES at different time steps (middle left); gauge-invariant contribution to qTDPES ($\epsilon_{wBO}$ in the figure) given in Eq.~\eqref{eq: bo tdpes in qEF} (middle right); geometric contribution to the qTDPES ($\epsilon_{kin}$ in the figure) given in Eq.~\eqref{eq: geo tdpes in qEF} (bottom left); gauge-dependent contribution to qTDPES ($\epsilon_{GD}$ in the figure) given in Eq.~\eqref{eq: gd tdpes in qEF} (bottom right).

For weak coupling, \ie $\lambda D=0.01$, for both $N_p\to \infty$ and $N_p=1$, near $q_r/q=0$, time-dependent nearly-singular ``spikes'' were found to originate mainly from $\TDPES^{\mathrm{qEF}}_{geo}$. Instead, $\TDPES^{\mathrm{qEF}}_{GD}$ is basically constant or slowly varying as a function of $q_r/q$ and $\TDPES^{\mathrm{qEF}}_{c}$ is nearly harmonic. The appearance of these spikes causes the marginal density to develop a multi-modal shape, with low-density regions localized at the positions of the peaks. In addition, the comparison between $\TDPES^{\mathrm{qEF}}_{c}$ and $\TDPES^{\mathrm{qEF}}_{geo}$ shows that both functions develop peaks a the same positions, even though in $\TDPES^{\mathrm{qEF}}_{c}$ the peaks do not tend to a singular behavior. For strong coupling, \ie $\lambda D=0.4$, and for $N_p=1$, the structure of the qTDPES is much more complex, showing multiple smooth peaks, with $\TDPES^{\mathrm{qEF}}_{geo}$ and $\TDPES^{\mathrm{qEF}}_{GD}$ of similar magnitude as $\TDPES^{\mathrm{qEF}}_{c}$. Overall, the strong features of $\TDPES^{\mathrm{qEF}}_{geo}$ are mainly related to the small (effective) mass associated to the photonic degrees of freedom, which make the geometric contribution to the qTDPES of comparable strength as the other contributions and, thus, of challenging to model in the context of trajectory-based techniques.

Similarly to the behavior of $\TDPES^{\mathrm{qEF}}_{geo}$ as function of $\phq$, the electronic $\phq,t$-dependent coefficients, namely
\be
\label{eq.expcofqBO}
C^{\mathrm{qEF}}_n(\phq,t)=\frac{\chi_n(\phq,t)}{\chi(\phq,t)}
\ee
present particularly interesting dynamics. Here, the index $n$ labels the $n$-th eigenfunction of the Hamiltonian $\hat H-\hat T_{p}$. In all cases reported in Fig.~\ref{fig:cavityhoff}, the coefficients $|C^{\mathrm{qEF}}_n(\phq,t)|^2$ show rich dynamics and vary strongly as functions of $\phq$ (top right plots in each panel of Fig.~\ref{fig:cavityhoff}). This behavior shows remarkable differences if compared to the electron-nuclear problem, where the corresponding expansion coefficients of the electronic conditional amplitude behave mainly like step functions, with the steps appearing in regions of low nuclear density. In the exact electron-nuclear factorization, extensive analyses were conducted based on model systems and semiclassical arguments, which provided the bases for the approximations used in the exact-factorization-based algorithms for nonadiabatic molecular dynamics~\cite{AgoAbeSuzMinMaiGro-JCP-15}. Unfortunately, these observations cannot be straightforwardly extended to the qEF case and more refined modeling is probably necessary to derive approximate schemes.

Comparison of panels b and c of Fig.~\ref{fig:cavityhoff} shows that the behavior of the system changes qualitatively when increasing the coupling strength, since from $\lambda D=0.01$ to $\lambda D=0.4$ the system moves towards the nonperturbative domain. Experimentally, this so-called ultrastrong coupling regime is typically defined by a normalized coupling strength $\eta=g/\omega_c>0.1$, where $g$ is related to the parameters of the model of Ref.~\cite{Maitra_EPJB2018} as $g=\lambda D \sqrt{\omega_c/2}$. In the case of the results in Fig.~\ref{fig:cavityhoff}c, $g=0.10$ and $\eta=0.45$. Values up to $\eta\approx0.3$ have already been achieved in systems using organic molecules (and even higher couplings are experimentally accessible, in \eg superconducting circuit platforms, see Ref.~\cite{FriskKockum2019} and references therein). The complexity of the photonic dynamics in the ultrastrong coupling regime, accompanied by strong deviations from the harmonic behavior, is particularly well captured by the exact factorization formalism. 

\begin{figure}
    \centering
    \includegraphics[width=\columnwidth]{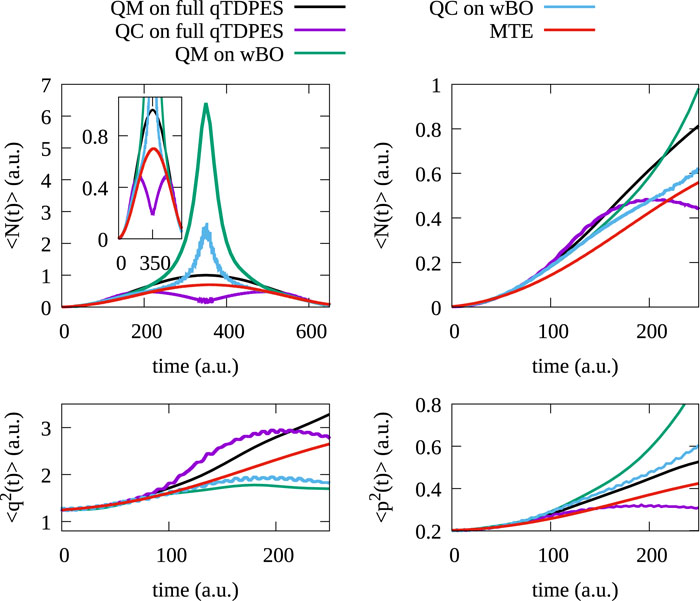}
    \caption{Top panels: average photon number as a function of time, during a Rabi period of the model simulated in Ref.~\cite{Maitra_JCP2022} (left), and during the first $~250$~a.u. (right). Bottom panels: average values of $\langle \hat q^2(t)\rangle$ (left) and of $\langle \hat p^2(t)\rangle$ (right) during the first $~250$~a.u. The various colors of the curves refer to the photon dynamics simulated with the support of the full qTDPES, quantum-mechanically (black curves) and quantum-classically (purple curves), and with support of $\TDPES_c^{\mathrm{qEF}}(\phq,t)$ (indicated as wBO in the figure), quantum-mechanically (green curves) and quantum-classically (light blue curves). The results obtained with MTE are also shown as reference (red curves).
    Reproduced from Rosenzweig, B., Hoffmann, N. M., Lacombe, L., \& Maitra, N. T. (2022). The Journal of Chemical Physics, 156(5), 054101, with the permission of AIP Publishing.}
    \label{fig: photon number}
\end{figure}

In the framework of qEF, Maitra and co-workers~\cite{Maitra_JCP2022} investigated the explicit application of trajectory-based methods to describe the dynamics of the photonic degrees of freedom in a model system similar to Ref.~\cite{Maitra_EPJB2018} in a similar parameter regime as Fig. \ref{fig:cavityhoff}b. To examine the importance of the components of the qTDPES and the applicability of typical trajectory-based techniques, the authors performed various kinds of dynamics simulations, from quantum to classical, using as driving potential either the full qTDPES of Eq.~\eqref{eq: tdpes in qEF} or the gauge-invariant contributions $\TDPES_c^{\mathrm{qEF}}$ of Eq.~\eqref{eq: bo tdpes in qEF} (indicated as wBO in Ref.~\cite{Maitra_JCP2022}). Figure~\ref{fig: photon number} shows the observables computed as expectation values of the number operator (upper panels), of the squared of the photon displacement (lower left panel) and of the squared of the photon momentum (lower right panel), as functions of time. The curves labeled QM refers to the quantum dynamics propagation whereas those labeled QC refer to the trajectory-based simulations. For the latter, the initial conditions are Wigner-sampled harmonically and the forces are determined either as $-\nabla \TDPES^{\mathrm{qEF}}(\phq,t)$ (QC on full qTDPES) or as $-\nabla \TDPES^{\mathrm{qEF}}_c(\phq,t)$ (QC on wBO). As reference, multi-trajectory Ehrenfest (MTE) results are also shown, where the photon-electronic coupling is treated in a mean-field manner.

The quantum dynamics result in Fig.~\ref{fig: photon number} (upper left) shows that, during the first 600~a.u. of dynamics, \ie a Rabi period in the model, one photon is emitted and reabsorbed. A reasonably accurate result is obtained with the quantum dynamics propagated using only $\TDPES^{\mathrm{qEF}}_c(\phq,t)$ until about 200~a.u. (see top right panel), and after that the average number of photons is largely overestimated until the half Rabi period (after that, the symmetric behavior is observed until the end of the period). The trajectories reproduce qualitatively well the behavior of the average photon number, even though quantitative agreement is lacking in both bases. Since the average photon number can be written as a combination of $\langle \hat q^2(t)\rangle$ and $\langle \hat p^2(t)\rangle$, the comparisons in the lower panels of Fig.~\ref{fig: photon number} show the contributions that mainly deviate from the reference (black curves) results until half a Rabi period. It is worth highlighting that MTE results agree extremely well with the quantum reference and reproduce accurately these expectation values, even though quantitative agreement is missing.

\subsection{Cavity exact factorization}\label{sec: cEF}
The qEF approach beyond the clamped-nuclei approximation has not been employed so far for practical applications, especially since a complete quantum mechanical treatment of the electron-nuclear subsystem is generally not feasible. Nonetheless, a viable strategy for practical trajectory-based simulations that generalizes nonadiabatic molecular dynamics algorithms is the combination of pEF and qEF, yielding what we defined as cEF. In cEF, the photon-nuclear degrees of freedom are treated on equal footing as marginal variables with the electrons only appearing in the conditional amplitude,
\begin{align}
\Psi(\elr,\nucR,\phq,t)=\chi(\nucR,\phq,t)\Phi(\elr,t;\nucR,\phq)
\end{align}
Such a formulation of the problem lends itself naturally to a trajectory-based treatment of the photon-nuclear dynamics, which is fully determined by the $\nucR,\phq$-dependent cTDPES
\begin{align}\label{eq: tdpes in cEF}
\TDPES^{\mathrm{qEF}}(\nucR,\phq,t)=\TDPES_{c}^{\mathrm{cEF}}(\nucR,\phq,t)+\TDPES^{\mathrm{cEF}}_{geo}(\nucR,\phq,t)+\TDPES^{\mathrm{cEF}}_{GD}(\nucR,\phq,t)
\end{align}
and cTDVP
\begin{align}\label{eq: tdvp in cEF}
    \TDVP_\Gamma(\nucR,\phq,t) = \braket{\Phi(t;\nucR,\phq)|-i\hbar\nabla_\Gamma \Phi(t;\nucR,\phq)}_{\elr}
\end{align}
Equations~\eqref{eq: tdpes in cEF} and~\eqref{eq: tdvp in cEF} are equivalent to Eqs.~\eqref{eq: tdpes in pEF} and~\eqref{eq: tdvp in pEF} derived using the pEF approach (or to Eqs.~\eqref{eq: tdpes in qEF} and~\eqref{eq: tdvp in qEF} derived using the qEF approach), and the index $\Gamma$ runs over the $N_n$ nuclear degrees of freedom and over the $N_p$ modes ($\Gamma = \nu,\alpha$). As in pEF and qEF, Eq.~(\ref{eq: tdpes in cEF}) is the sum of three contributions, namely
\begin{align}
    \TDPES_{c}^{\mathrm{cEF}}(\nucR,\phq,t)&=\braket{\Phi|\hat H - \hat T_n-\hat T_p|\Phi}_{\elr}\label{eq: bo tdpes in cEF}\\
    \TDPES^{\mathrm{cEF}}_{geo}(\nucR,\phq,t)&= \sum_\Gamma \frac{\hbar^2\langle \nabla_\Gamma \Phi|\nabla_\Gamma \Phi\rangle_{\elr}}{2M_\Gamma}-\frac{[\TDVP^{\mathrm{cEF}}_\Gamma(\nucR,\phq,t)]^2}{2M_\Gamma}\label{eq: geo tdpes in cEF}\\
    \TDPES^{\mathrm{cEF}}_{GD}(\nucR,\phq,t)&=\braket{\Phi|-i\hbar \partial_t|\Phi}_{\elr}\label{eq: gd tdpes in qcEF}
\end{align}
with $M_\Gamma = \lbrace M_\nu,M_\alpha\rbrace =  \lbrace M_\nu,1\rbrace$. Using the cEF approach in combination to an approximate scheme to solve the PEN dynamics allows one to circumvent the calculation of PoPES and related quantities, like couplings, in a vector space of larger dimension than the electronic vector space in cavity-free conditions. In addition, as in qEF, working with the cEF approach does not require truncations of the photon states at a maximal photon number, and allows -- in principle -- for the inclusion of an arbitrary number of photons, significantly broadening the potential applications and efficiency~\cite{Kowalewski2016}. With this motivation, Maitra~\cite{Maitra_JCP2022} and Agostini~\cite{Agostini_JCP2024_2} analyzed the cEF perspective.

The cEF approach has been studied in Ref.~\cite{Agostini_JCP2024_2} in a situation that somehow combines the features of the models presented in Secs.~\ref{sec: pEF} and~\ref{sec: qEF}. There, the simulated model represents a two-electronic-state one-dimensional (in nuclear space) system which is strongly coupled to a single cavity mode. The cavity-free BOPESs are two identical parabolas, either shifted in position or shifted in energy. When the system is strongly coupled to the cavity, the former configuration of the BOPESs yields nonadiabatic dynamics reminiscent of Ref.~\cite{Maitra_PRL2019}, when the cavity is in resonance to energy gap between the two electronic states at the Franck-Condon point, whereas the latter configuration yields Rabi oscillations between the two electronic states similarly to Ref.~\cite{Maitra_EPJB2018}, when the cavity is in resonance with the energy gap separating the two parallel BOPESs.

\begin{figure}
    \centering
    \includegraphics[width=1\linewidth]{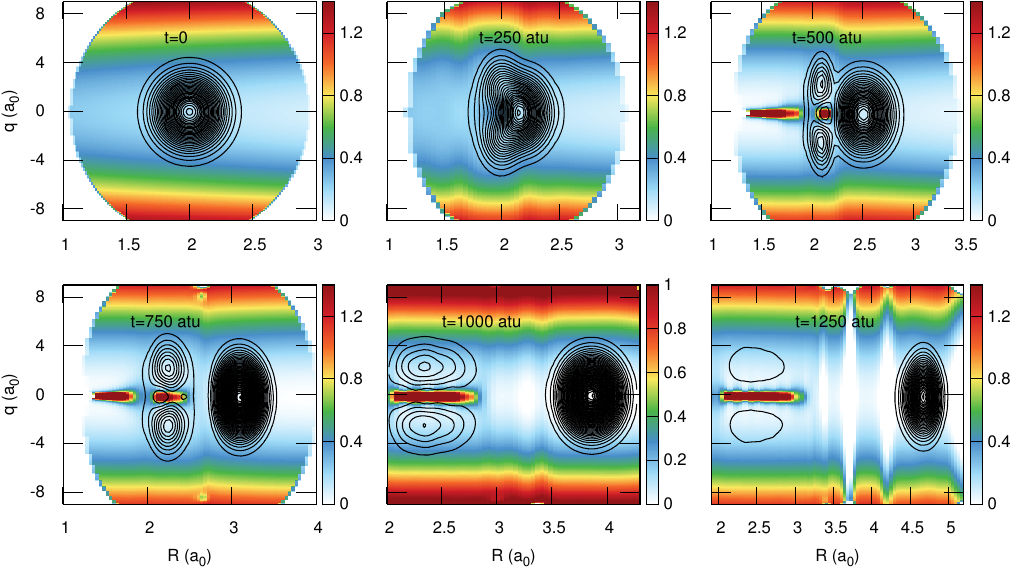}
    \caption{Color map of the cTDPES for the nonadiabatic model of Ref.~\cite{Agostini_JCP2024_2} at the times indicated in the panels. The color bar is given in hartree. The nuclear density is superimposed at the same times as black contour lines. Reproduced from Sangiogo Gil, E., Lauvergnat, D, \& Agostini, F. (2024). The Journal of Chemical Physics, 161(8), 084112, with the permission of AIP Publishing.}
    \label{fig: tdpes in cEF}
\end{figure}
Figure~\ref{fig: tdpes in cEF} shows the cTDPES in the nonadiabatic situation, at different time steps along the dynamics, as color map with the superimposed black contour lines indicating the photon-nuclear density. Overall, the cTDPES has a shape that confines the marginal density along the photonic direction $q$ in the plots and that is somehow shallow along the nuclear direction $R$. In this way, the cTDPES allows a portion of the marginal density to move from the left to the right along $R$, while it confines in the Franck-Condon region the remaining portion of the density. This effect is achieved by modulating the shape of the cTDPES already at early times, \ie $t=250$~a.u., and by developing a barrier between $R=3$~a.u. and $R=3.5$~a.u. which is easily observable at $t=1000$~a.u. It is interesting to notice that along $q$, in the Franck-Condon region ($R=2$~a.u.), a localized peak at $q=0$ starts developing at $t=500$~a.u., suggesting that first excited state of the harmonic oscillator, with a node at the origin, is forming, related to the emission of photons.

\begin{figure}
    \centering
    \includegraphics[width=1\linewidth]{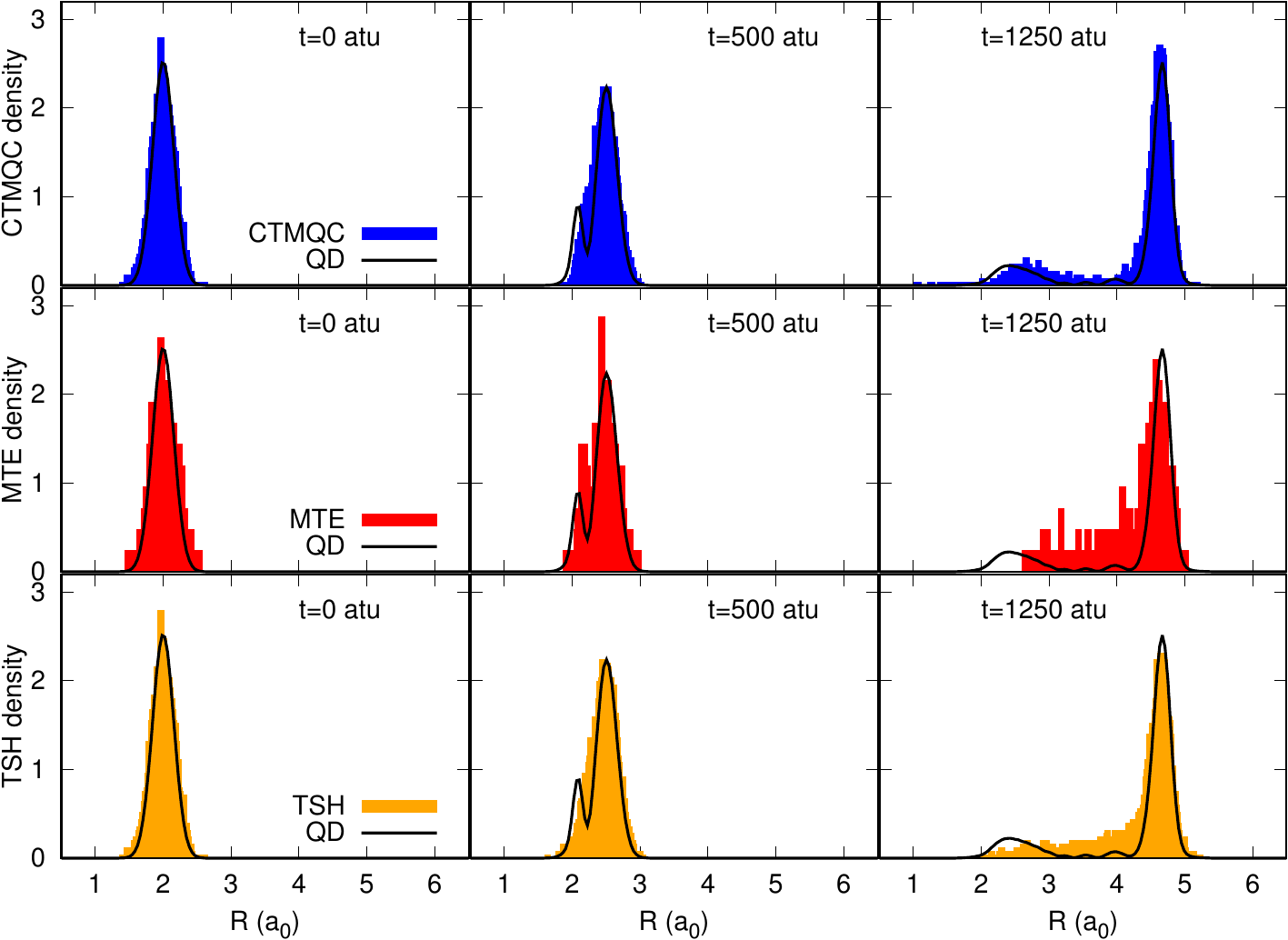}
    \caption{Histograms representing the distributions of trajectories to reproduce the marginal nuclear densities for the nonadiabatic model of Ref.~\cite{Agostini_JCP2024_2} calculated with CTMQC (blue), MTE (red), and TSH (orange) at the times indicated in the panel; the QD  marginal nuclear density is shown as black lines at the same times. Reproduced from Sangiogo Gil, E., Lauvergnat, D, \& Agostini, F. (2024). The Journal of Chemical Physics, 161(8), 084112, with the permission of AIP Publishing.}
    \label{fig: MQC in cEF}
\end{figure}

In Ref.~\cite{Agostini_JCP2024_2}, trajectory-based simulations were performed to approximate the dynamics just described, aiming to capture the coupled PEN dynamics using various nonadiabatic molecular dynamics methods. Specifically, three schemes were applied: the coupled-trajectory mixed quantum classical (CTMQC, shown in blue in the figures below) scheme derived from the exact factorization~\cite{Gross_PRL2015, Agostini_PCCP2024}, multi-trajectory Ehrenfest (MTE, shown in red in the figures below) and Tully surface hopping (TSH, shown in orange in the figures below)~\cite{Tully_JCP1990}. Overall, the dynamics simulated with the three schemes is quite similar, which can be observed at various time steps, by comparing the quantum marginal nuclear density (black lines) to the histograms constructed from the distributions of trajectories (CTMQC in the top row, MTE in the middle row, TSH in the bottom row) in Fig.~\ref{fig: MQC in cEF}. Nonetheless, it is clear that, even if not with perfect agreement with quantum dynamics, CTMQC is able to reproduce better the splitting of the marginal nuclear density observed at long times.

\begin{figure}[hbt!]
    \centering
    \includegraphics[width=1\linewidth]{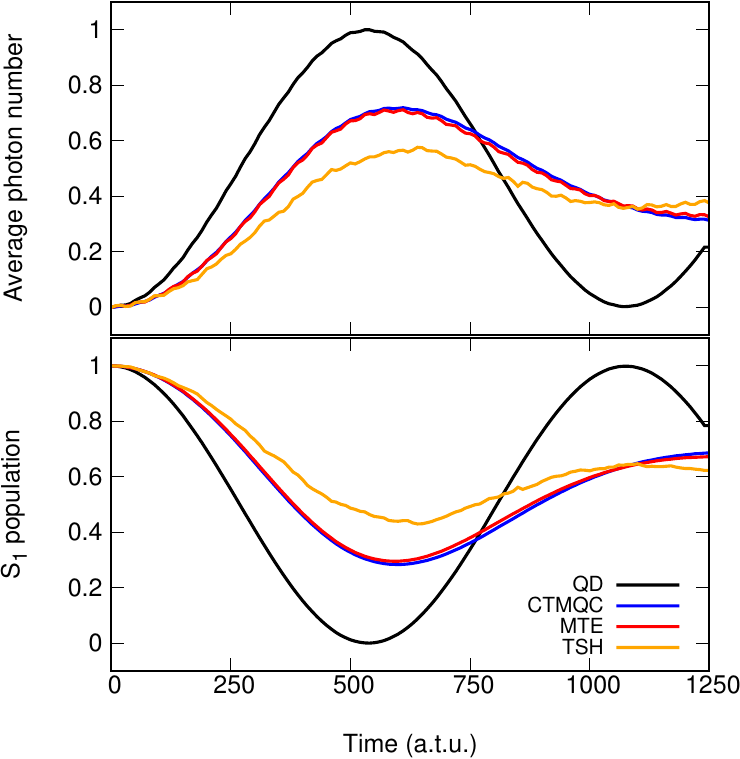}
    \caption{Average photon number as function of time (top panel) and population of the excited state as function of time (bottom panel) for the nonadiabatic model of Ref.~\cite{Agostini_JCP2024_2}. Quantum dynamics results are shown in black and three different trajectory-based approaches are compared: CTMQC in blue, MTE in red and TSH in orange. Reproduced from Sangiogo Gil, E., Lauvergnat, D., \& Agostini, F. (2024). The Journal of Chemical Physics, 161(8), 084112, with the permission of AIP Publishing.}
    \label{fig: average photon number}
\end{figure}

\begin{figure}
    \centering
    \includegraphics[width=1\linewidth]{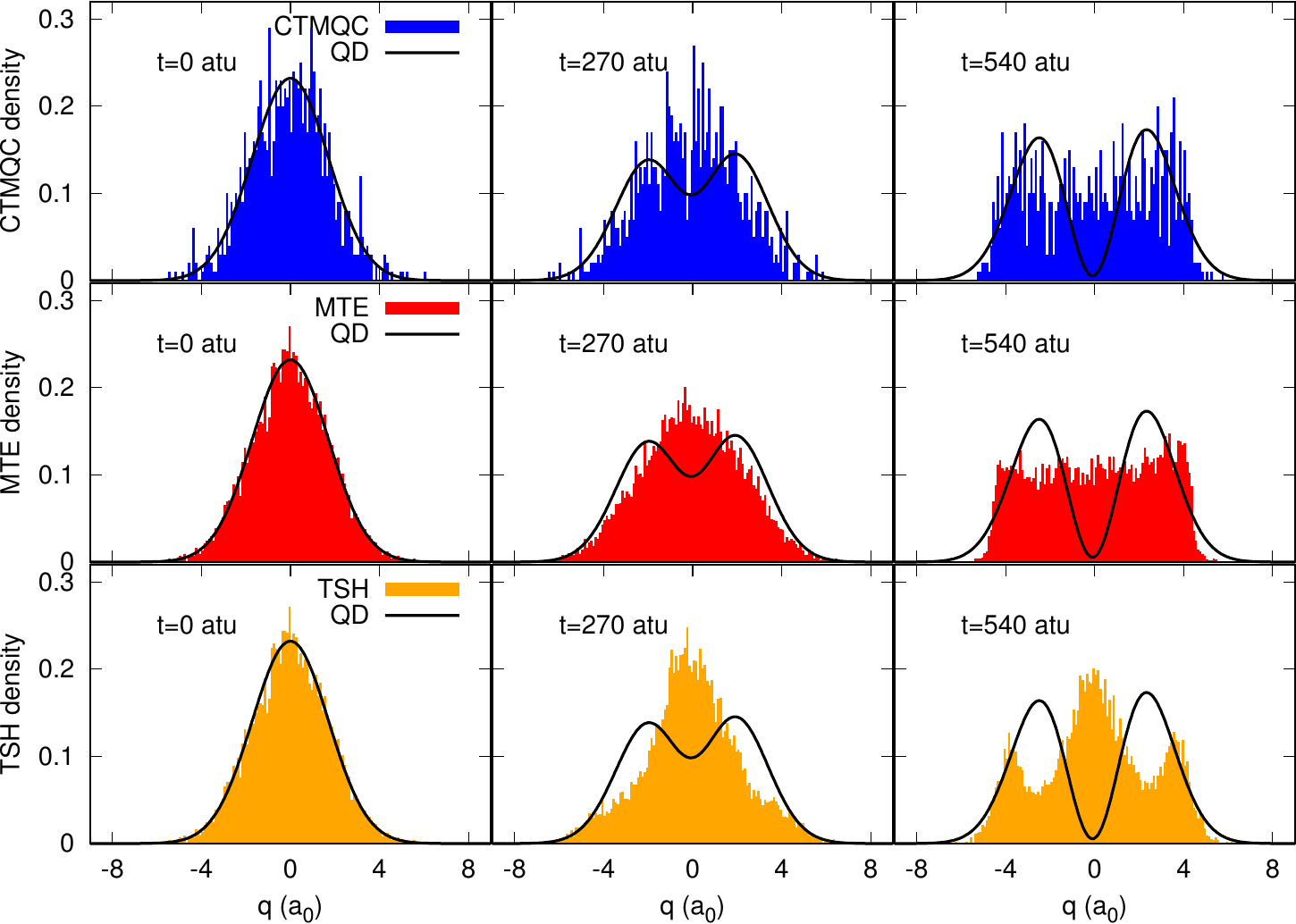}
    \caption{Same as Fig.~\ref{fig: MQC in cEF} but for the Rabi model of Ref.~\cite{Agostini_JCP2024_2}. Reproduced from Sangiogo Gil, E., Lauvergnat, D., \& Agostini, F. (2024). The Journal of Chemical Physics, 161(8), 084112, with the permission of AIP Publishing.}
    \label{fig: marg dens q}
\end{figure}

Figures~\ref{fig: average photon number} and~\ref{fig: marg dens q} report some results in the Rabi situation, namely the average photon number as function of time (top panel of Fig.~\ref{fig: average photon number}) along with the time trace of the population of the excited state (bottom panel of Fig.~\ref{fig: average photon number}), and the marginal photonic density (in Fig.~\ref{fig: marg dens q}). The quantum dynamics results (black lines) show that the photon emission is accompanied by the population decay from the excited state to the ground state, and when the average photon number reaches unity the emitted photon is reabsorbed to bring back the electronic population fully in the excited state.
It was observed in Ref.~\cite{Agostini_JCP2024_2} that in this case the photon-nuclear density remains localized around the initial positions $q=0$ and $R=2$~a.u., and the dynamics does not manifest any peculiar behavior along the $R$ direction. Nonetheless, the dynamics along $q$ is quite interesting, similarly to Refs.~\cite{Maitra_EPJB2018}, as the cavity switches between the zero-photon state and the one-photon state. To achieve this, the density along $q$ needs to evolve between the ground state and the first excited state of the harmonic oscillator. As observed in Sec.~\ref{sec: qEF}, in order to produce this effect the cTDPES, and specifically its geometric component, needs to develop a very localized peak at $q=0$, for all values of $R$. 

Since $\TDPES_{geo}^{\mathrm{cEF}}$ is neglected in CTMQC, despite this algorithm being derived from the exact factorization, one observes disagreement between CTMQC (in blue in Fig.~\ref{fig: marg dens q}) and quantum dynamics results. MTE and TSH do not perform better than CTMQC, actually TSH provides a poorer description of the dynamics than MTE and CTMQC. 
Overall, the trajectory-based methods fail in reproducing the correct splitting of the density, thus underestimating the photon-emission process, even though at the final time reported in the figure, CTMQC shows the lowest presence of trajectories at $q=0$, probably due to the presence of a small barrier in $\TDPES_{c}^{\mathrm{cEF}}$ similar to the barrier observed in $\TDPES_{c}^{\mathrm{qEF}}$ in Fig.~\ref{fig:cavityhoff} (in all middle right panels).

\subsection{Exact electron factorization in cavity quantum electrodynamics} \label{sec: ceEF}
Beyond the time-dependent studies presented above, an alternative application of exact factorization in cQED has been proposed in the literature by Tokatly and co-workers~\cite{Tokatly_EPJB2018}, by focusing on the calculation of the electronic marginal eigenfunctions in the regime of strong coupling to a quantized light field. Such a framework shows connections between the EEF presented in Sec.~\ref{sec:EEF} and the inverse exact factorization developed in a time-dependent situation in Refs.~\cite{Suzuki_PRA2014,Khosravi2015,Khosravi2017}. In Table ~\ref{tab:typesofCavityEF} we refer to this as cavity electron exact factorization (ceEF).

In this case, the Hamiltonian describing the photon-electron system is $\hat H(\elr,\phq) = \hat T_{el}+\hat V_{el}(\elr)+\hat T_p+\hat V_p(\phq)+\hat H_{el,ph}(\elr,\phq)$ with $\hat T_{el}$ the electronic kinetic energy and $\hat V_{el}(\elr)$ the interaction among the electronic degrees of freedom (and eventually the effect of an external potential). Using the time-independent formalism presented in Sec.~\ref{sec: theory}, the $n$-th eigenstate of the photon-electron stationary problem is written as
\begin{align}
    \Psi_n(\elr,\phq) = \chi_n(\elr)\Phi_n(\phq;\elr)
\end{align}
An equation analogous to Eq.~\eqref{eq:conditionalMain stationary}, with $\hat H_c(\elr,\phq) = \hat V_{el}(\elr)+\hat T_p+\hat V_p(\phq)+\hat H_{el,ph}(\elr,\phq) = \hat H(\elr,\phq) - \hat T_{el}$ yields the gauge-independent exact potential energy surface $\varepsilon_n(\elr)$ associated to each one of the $n$ eigenstates. The exact potential energy surface reads
\begin{align}
    \varepsilon^{(n)}(\elr) =\varepsilon^{(n)}_c(\elr)+\varepsilon^{(n)}_{geo}(\elr)
\end{align}
which is the sum of two terms, namely
\begin{align}
   \varepsilon^{(n)}_c(\elr) &= \langle\Phi_n(\elr)| \hat H-\hat T_{el}| \Phi_n(\elr)\rangle_{\phq} \\
   \varepsilon^{(n)}_{geo}(\elr) &= \sum_i \frac{\hbar^2 \langle\nabla_i\Phi_n(\elr)|\nabla_i\Phi_n(\elr)\rangle_{\phq}}{2m} -\frac{[\TDVP_i^{(n)}(\elr)]^2}{2m}
\end{align}
Here, the sum over $i$ runs over the electrons, $\nabla_i$ indicates the gradient with respect to the position of the $i$-th electron and $m$ is the electronic mass. The exact vector potential is
\begin{align}
    \TDVP_i^{(n)}(\elr) = \braket{\Phi_n(\elr)|-i\hbar \nabla_i\Phi_n(\elr)}_{\phq}
\end{align}
The exact potential energy surface and vector potentials yield the marginal electronic amplitude $\chi_n(\elr)$ via an effective stationary Schr\"odinger equation analogous to Eq.~\eqref{eq:marginalMain stationary}.

\begin{figure}
    \centering
    \includegraphics[width=1\linewidth]{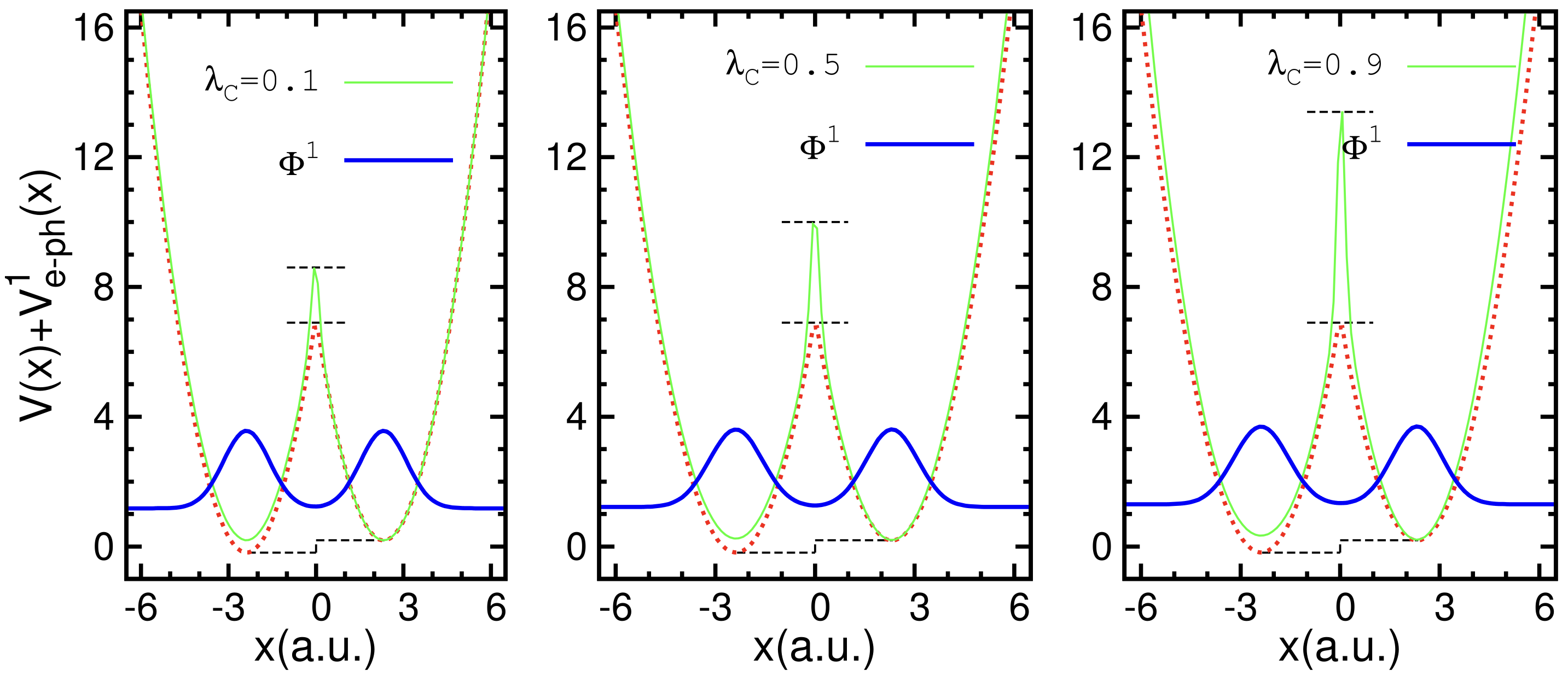}
    \caption{Comparison between the original double-well potential of the model studied in Ref.~\cite{Tokatly_EPJB2018} (dashed red curves) and the exact potential energy surface (continuous green curves) yielding the electronic marginal amplitude associated to the the first excited state. The marginal density is shown in blue. The three panels refer to increasing values of the photon-electron coupling from left to right. Reproduced from Abedi, A., Khosravi, E., \& Tokatly, I. V. (2018). The European Physics Journal B, 91, 194, with the permission of Springer Nature.}
    \label{fig: el 1st eigenstate in cQED}
\end{figure}

In Ref.~\cite{Tokatly_EPJB2018} the authors investigated the effect of the strong photon-electron coupling on the eigenstates of an asymmetric double well potential. The model features a single electron under the effect of an external potential in the form of two identical parabolas shifted in position and in energy. The parameters of the model were chosen such that in cavity-free conditions the ground state is localized in a well and the excited state is localized in the other well. If the frequency of the cavity mode is chosen such that the two lowest electronic states are brought into resonance, the electronic excited state becomes completely delocalized in the two wells, independently of the coupling strength. This effect can be clearly appreciated in Fig.~\ref{fig: el 1st eigenstate in cQED} which compares the original asymmetric double well potential (dashed red curves) to the exact potential energy surface (continuous green curves) calculated in a gauge that sets to zero the exact vector potential. In the panels of Fig.~\ref{fig: el 1st eigenstate in cQED} the $x$-axis indicates the position of the marginal particle, \ie the electron, and the blue curves show the marginal probability density ($\chi_1(\elr)$ in our notation). The three panels differ in the value of the coupling strength between the electronic excitation and the cavity mode, which increases from left to right. The exact potential energy surface for the first excited marginal amplitude features various difference if compared to the original double well potential: the two wells are brought at the same level, yielding in this way a symmetric double well; the barrier between the two wells increases such that the delocalization of the electronic densities on both sides is stabilized; as the coupling constant increases, the potential well squeezes having the effect of localization of the density in each well, that was referred to as polaritonic squeezing of the excited state.

Beyond the observations of the features of the exact potential energy surface and of the exact marginal density, Ref.~\cite{Tokatly_EPJB2018} also developed an approximate procedure to determine these quantities without resorting to the exact solution of the full coupled problem.

\section{Conclusions}\label{sec:conclusions}
We reviewed some applications of the exact factorization aiming to provide original perspectives and novel ideas for numerical schemes to solve the quantum-mechanical many-body problem in various situations. Specifically, we reported on two quite different physical situations, namely the stationary many-electron interacting problem and the time-dependent molecular problem in cavity quantum electrodynamics, thus demonstrating the flexibility of the exact factorization formalism. To tackle these kinds of problems, we reported on the effort of many research groups that developed, on the one hand, the exact electron-only factorization in connection with density-functional theory, and, on the other hand, the exact photon-electron-nuclear factorization mainly in connection with (nonadiabatic) molecular dynamics. 

Beyond its role as a theoretical tool for the analysis of the exact many-body problem in some typical model situations, the exact factorization formalism has also served as a starting point for developing systematic and physically motivated approximations. Exact-factorization-based approaches demonstrated to have the capabilities to achieve the efficiency and the accuracy of traditional methods, and have the potential to outperform standard techniques thanks to the rigorous theoretical construction allowing one to naturally recognize and include missing contributions.

\section*{Data availability statement}
Data sharing is not applicable to this article as no new data were created or analyzed in this study. 


\section*{Acknowledgements}
This project has received funding from the European Union's Horizon 2020 research and innovation programme under the Marie Sk{\l}odowska-Curie grant agreement No 101104947 and from the French Agence Nationale de la Recherche via the project STROM (Grant No. ANR-23-ERCC-0002).



\end{document}